\documentclass{article}
\usepackage[margin=3cm]{geometry}

\usepackage{amsmath,amssymb,amsfonts,mathtools,xfrac}
\usepackage{mathrsfs} 
\usepackage{cancel}
\usepackage[labelformat=simple]{subcaption}

\DeclareCaptionLabelFormat{subcaptionlabel}{\normalfont(\textbf{#2}\normalfont)}
\usepackage{float}

\usepackage{tikz}
\usetikzlibrary{matrix}
\usepackage{setspace}

\usepackage{graphicx,color}
\usepackage{accents}

\usepackage{cite}

\usepackage[normalem]{ulem}

\newcommand{\ourG}{\mathbf{G}}
\newcommand{\ourB}{\mathbf{B}}

\numberwithin{equation}{section}

\usepackage{hyperref}
\hypersetup{colorlinks=true,allcolors=blue,urlcolor=cyan}

\title{Spatial curvature in coincident gauge $f(Q)$ cosmology
	}
\author{
  Erik Jensko\footnote{Email: erik.jensko@ucl.ac.uk} \\
  Department of Mathematics, University College London, \\
  Gower Street, London WC1E 6BT, United Kingdom}

\date{\today} 

\begin{document}

\maketitle

\begin{abstract}
In this work we study the Friedmann-Lema\^{i}tre-Robertson-Walker cosmologies with arbitrary spatial curvature for the symmetric teleparallel theories of gravity, giving the first presentation of their coincident gauge form.
Our approach explicitly starts with the cosmological Killing vectors and constructs the coincident gauge coordinates adapted to these Killing vectors. We then obtain three distinct spatially flat branches and a single spatially curved branch.
Contrary to some previous claims, we show that all branches can be studied in this gauge-fixed formalism, which offers certain conceptual advantages. We also identify common flaws that have appeared in the literature regarding the coincident gauge.
Using this approach, we find that both the flat and spatially curved solutions in $f(Q)$ gravity can be seen as equivalent to the metric teleparallel $f(T)$ models, demonstrating a deeper connection between these theories. This is accomplished by studying the connection equation of motion, which can be interpreted as a consistency condition in the gauge-fixed approach. Finally, we discuss the role of diffeomorphism invariance and local Lorentz invariance in these geometric modifications of gravity.

\end{abstract}

\clearpage

\setcounter{tocdepth}{2}
\tableofcontents

\section{Introduction} 

Modifications and extensions of General Relativity (GR) take many different forms, with a rich and varied history~\cite{Goenner:2004se}. These modifications have been motivated for a number of reasons, but recently a strong motivation comes from the growing observational tensions in cosmology~\cite{Perivolaropoulos:2021jda}. The Hubble tension and $\sigma_{8}$ tension are both concerned with the discrepancy between early and late-time cosmological observations, and has so far been difficult to solve within the standard $\Lambda$CDM framework~\cite{Verde:2019ivm,DiValentino:2020vvd}. As such, cosmology is a perfect testing ground for beyond-GR theories, which may be able to alleviate these tensions. We refer to~\cite{Abdalla:2022yfr} for an overview of these discordances and possible solutions. 
 
A common theme in modified gravity theories is an explicit introduction of new degrees of freedom, such as new scalar, vector, or tensor fields. Moreover, almost all alternative theories can be reformulated in this way (e.g., rewriting $f(R)$ gravity as a scalar-tensor theory, or Lorentz violations in terms of additional background fields~\cite{Clifton:2011jh,Nojiri:2010wj,Nojiri:2017ncd}). A more fundamental departure is to reconsider the Levi-Civita geometry characterised by a torsion-free and metric-compatible connection. 
Dropping these assumptions introduces new geometric components related to the affine connection, which naturally arises in a gauge formulation of gravity~\cite{Hehl:1994ue,CANTATA:2021ktz}. One such framework is \textit{teleparallel geometry}, which has vanishing curvature but non-vanishing torsion and non-metricity. If torsion also vanishes we are led to \textit{symmetric teleparallelism}, characterised by a non-metric connection. Instead, we can have a metric-compatible connection with torsion only, known as \textit{metric teleparallelism}. Again, these geometries have a natural gauge-theoretic description~\cite{Blagojevic:2013xpa}, further motivating their study.

Within these geometries there exists equivalent formulations of General Relativity, known as the teleparallel equivalents of General Relativity (TEGR)~\cite{Nester:1998mp,Adak:2005cd,Aldrovandi:2013wha,deAndrade:2000kr}. These serve as starting points for the modifications $f(T)$ and $f(Q)$ gravity~\cite{Ferraro:2006jd,Bengochea:2008gz,BeltranJimenez:2017tkd,Bahamonde:2021gfp}, where $T$ and $Q$ are the torsion and non-metricity scalars respectively. For these non-linear modifications, the teleparallel connection becomes dynamical and has its own equations of motion. Hence, these include genuine new degrees of freedom, though isolating them has proven difficult~\cite{Blixt:2020ekl,DAmbrosio:2023asf}. Because both theories include models that reduce to GR within some specific regimes, they have shown promise at constructing viable alternatives to the $\Lambda$CDM paradigm~\cite{Khyllep:2022spx,Bahamonde:2021gfp,Xu:2018npu}. However, there is still work to be done on the theoretical aspects of these theories. For instance, $f(T)$ gravity suffers a strong coupling problem, indicated by the absence of dynamical modes around Minkowski and maximally symmetric backgrounds (up to at least second-order)~\cite{BeltranJimenez:2020fvy,Hu:2023juh}. Hence, the predictions of perturbative calculations cannot be trusted, even at arbitrarily low energies. Similar pathologies appear generically in $f(Q)$ gravity around spatially flat cosmological backgrounds, where modes are either strongly coupled or ghost instabilities arise~\cite{BeltranJimenez:2019tme,Gomes:2023tur,Zhao:2024kri}. A deep understanding of these geometric frameworks is therefore of paramount importance, especially around more general cosmological backgrounds.

A key property of the teleparallel theories is that they possess a special gauge where the connection trivialises and vanishes. In the symmetric teleparallel geometry it is known as the \textit{coincident gauge}~\cite{Nester:1998mp,Adak:2005cd}, while in the metric teleparallel geometry it is the \textit{Weitzenb\"{o}ck gauge}~\cite{Aldrovandi:2013wha}. These gauges correspond to specific coordinate systems and tetrad frames respectively; hence, one cannot arbitrarily choose the form of the metric (or tetrad) while working in these gauges~\cite{Adak:2005cd,Krssak:2015oua}. Moreover, as these are fundamentally just gauge choices, they should not affect the physics nor the degrees of freedom within these theories. Parallels can be made with the unitary gauge of massive gravity~\cite{deRham:2014zqa}, and similarly the affine connection of the teleparallel theories can be interpreted as the Stueckelberg fields of broken symmetries~\cite{BeltranJimenez:2019tme} --- a point we will return to shortly.

In the symmetric teleparallel geometry, the existence of the coincident gauge has long been known\footnote{In 1927, Eisenhart~\cite{eisenhart1927non} gives a simple and explicit proof of the existence of these coincident gauge coordinates (where the affine connection vanishes globally) for geometries with a flat and symmetric connection.}, but only recently has it been applied to gravitational theories and cosmology~\cite{Nester:1998mp,Adak:2008gd}. Now it is also being utilised in the modified $f(Q)$ gravity theories~\cite{Bahamonde:2022zgj,BeltranJimenez:2019tme}. This has usually been in the case of flat Friedmann-Lema\^{i}tre-Robertson-Walker (FLRW) cosmology in Cartesian coordinates, which is known to be consistent with the coincident gauge~\cite{BeltranJimenez:2019tme}. We also showed this to be true for (exterior) spherically symmetric spacetimes in isotropic Cartesian coordinates~\cite{Boehmer:2021aji}. More recently, this was extended to more general spherically symmetric spacetimes~\cite{Bahamonde:2022zgj}, where the coordinate systems are highly non-trivial and multiple different branches of solution exist. Alternatively, one can work in an arbitrary gauge that is better suited to the symmetries of a given spacetime, but the connection will have non-vanishing components, see for instance~\cite{DAmbrosio:2021zpm,Zhao:2021zab,Lin:2021uqa}.
 
 On the other hand, assuming the vanishing of the affine connection can lead to inconsistencies if the coordinates do not in fact satisfy the coincident gauge, such as in~\cite{Mustafa:2021ykn,Silva:2024kix,Junior:2023qaq,Rastgoo:2024udl,Nashed:2024jmw,Maurya:2024dwn,Kirezli:2024xxj,Nashed:2024jqw,Maurya:2024wtj,Chaudharya:2024zvc,Maurya:2022wwa,Maurya:2023muz,Mustafa:2023kqt,Parsaei:2022wnu,Mohanty:2024pzh,Ditta:2023xhx,Mandal:2021qhx,Mustafa:2021bfs,Kaur:2024biz,ZeeshanGul:2024oui,Hassan:2021egb,Bhattacharjee:2024xkq,Chalavadi:2023sif,Junior:2024xmm,Wang:2021zaz,Sahoo:2023dus,Errehymy:2022gws,Hassan:2021vzk,Sharma:2021egn,Ditta:2023huk}. In such cases, any resulting analyses will necessarily be flawed unless a non-vanishing affine connection is also taken into account.
A common side effect is that the dynamical equations force $f''(Q)=0$, reducing the model to the symmetric teleparallel equivalent of General Relativity\footnote{When $f(Q) = \alpha Q + \beta $, the theory reduces to standard GR with a cosmological constant, which has an `enhanced' diffeomorphism symmetry due to the connection becoming non-dynamical. In that case, the coincident gauge has no requirements because the connection plays no physical role in the theory.}. Because of this, it has sometimes been argued that this gauge choice precludes certain solutions, or is incompatible with non-trivial $f(Q)$ models for certain spacetimes altogether\footnote{Note that some authors define or interpret the coincident gauge as that in which the connection equation of motion is satisfied identically~\cite{Paliathanasis:2023pqp,Paliathanasis:2023hqq,Krssak:2024xeh}. This is a subtle point, as it does vanish in some simple cases, but is not a generic feature of the coincident gauge, as shown in Section~\ref{sec:fe}. Alternatively, some authors correctly point out that the coincident gauge is not consistent with a given metric \textit{fixed to a set of coordinates}, and this is simply a case of trying to fix two incompatible gauges.}~\cite{Lohakare:2023ugg,Junior:2024xmm,Shi:2023kvu,Shabani:2023nvm,Paliathanasis:2023raj,Maurya:2024dwn,Wang:2021zaz,Nashed:2024jqw,Mohanty:2024pzh,DAmbrosio:2021zpm}. We will show this not to be the case, though it is often more difficult to work in these coordinates. This fact appears to not be widely known, but good explanations can be found in~\cite{BeltranJimenez:2018vdo,Heisenberg:2023lru,Bello-Morales:2024vqk,Gomes:2023hyk}. We hope to further elucidate this topic here, providing systematic techniques to study these more complicated spacetimes in the coincident gauge.

In this work, we investigate the most general FLRW spacetimes in the symmetric teleparallel geometry. Many recent studies have shown that spatial curvature is not ruled out by observation~\cite{Handley:2019tkm,DiValentino:2019qzk,Vagnozzi:2020rcz,Vagnozzi:2020dfn,Dhawan:2021mel,Yang:2022kho}, and it has begun to see interest in these symmetric teleparallel theories too~\cite{Hohmann:2021ast,Guzman:2024cwa,DAmbrosio:2021pnd,Heisenberg:2022mbo,Dimakis:2022rkd,Subramaniam:2023okn,Shabani:2023xfn,Paliathanasis:2023kqs,De:2022jvo,Paliathanasis:2023pqp}.
 To be concrete, we focus on the $f(Q)$ models when deriving the field equations, but our analysis can be applied to any class of theories in the symmetric teleparallel framework. 
 Here we present the first study in the coincident gauge, working directly with the cosmological Killing vectors. We then proceed to study the dynamical equations of motion and constraint equations, showing how our results align with past works. It should be stressed that transforming to the coincident gauge does not change the physical degrees of freedom of a theory, but is useful for the identification of these degrees of freedom and for finding new solutions.
 
In the gauge-fixed approach taken here, the usual connection field equation instead arises as a consistency condition; remarkably, treating this as an off-shell constraint (independent of the choice of model $f$) leads to a direct equivalence with the $f(T)$ theories, which we discuss in detail. Such an equivalence between the $f(T)$ and $f(Q)$ theories has been noted for the `Cartesian branch' of spatially flat FLRW cosmologies \cite{DAmbrosio:2021pnd}, but not for any of the other solutions. We extend the analysis to the other branches of solution, including the spatially curved solutions. In fact, we show that this off-shell consistency condition implies that \textit{all} $f(Q)$ cosmologies are equivalent to $f(T)$ cosmologies, given some additional assumptions about model-independence. This result gives further indications that the two teleparallel frameworks are intricately related.

A key point that we focus on is the complete equivalence between the gauge-fixed formalism and the usual covariant formalism. This is a necessity for the mathematical validity of the theory, as it has long been known that the coincident gauge can always be found for any symmetric teleparallel theories~\cite{eisenhart1927non}. Therefore, if it were the case that fixing the coincident gauge gave different solutions, the theory would be inconsistent.
This is especially salient to show for the curved FLRW spacetimes,
 which have sometimes been claimed to be incompatible with the coincident gauge~\cite{Shabani:2023nvm,Shi:2023kvu}. The degrees of freedom that would usually reside in the teleparallel connection are instead transferred to the metric itself via the breaking of coordinate invariance. To make this clear, we lay out the gauge-fixed formulation of $f(Q)$ gravity, which can be seen as a diffeomorphism-breaking modification of the first-order Einstein action~\cite{Boehmer:2021aji,Jensko:2023lmn}. The equivalence with the covariant $f(Q)$ theories and the compatibility of the coincident gauge is then shown explicitly.

In Section~\ref{sec:teleparallel} we begin with the metric-affine framework and introduce the teleparallel theories and their modifications. The gauge-fixed formalism is then explained, showing the equivalence with the covariant theories at the level of equations of motion. In Section~\ref{sec:kvf} we study the FLRW Killing vector fields and the associated coincident gauge conditions. The general transformation to this gauge is found, resulting in three spatially flat branches and one spatially curved branch. The FLRW metric for all of these branches is then derived. We emphasise that the results of this section are completely general and apply to the geometric framework itself, and not any particular class of theory. Hence, these results also apply to the recent \textit{ghost-free} constructions in symmetric teleparallel geometries~\cite{Gomes:2023tur,Bello-Morales:2024vqk}. In Section~\ref{sec:fe} we return to the concrete example of $f(Q)$ gravity, finding the corresponding field equations and constraint equations. For the $k=0$ and $k<0$ cosmologies, we show that this reduces to the $f(T)$ gravity solution if the constraint equation is made to vanish off-shell for all choices of model $f$. Finally, in Section~\ref{sec:conc} we summarise our results.

\section{Teleparallel gravity} \label{sec:teleparallel}
A metric-affine geometry $(\mathcal{M},g,\bar{\Gamma})$ introduces a Lorentzian metric $g_{\mu \nu}$, which we take to have mostly plus signature, and independent affine connection $\bar{\Gamma}^{\lambda}_{\mu \nu}$ with covariant derivative $\bar{\nabla}$. This structure naturally gives rise to the notions of Riemann curvature, torsion, and non-metricity
\begin{align} \label{Riemann}
  \bar{R}_{\mu \nu \lambda}{}^{\rho} &= 2 \partial_{[\mu} \bar{\Gamma}{}^{\rho}_{\nu] \lambda} +
  2 \bar{\Gamma}{}^{\rho}_{[\mu | \sigma} \bar{\Gamma}{}^{\sigma}_{\nu] \lambda} \,, \\
    T^{\lambda}{}_{\mu \nu} &= 2 \bar{\Gamma}^{\lambda}_{[\mu \nu]} =
  \bar{\Gamma}^{\lambda}_{\mu \nu} - \bar{\Gamma}^{\lambda}_{\nu \mu} \,, \\
  Q_{\lambda \mu \nu} &= - \bar{\nabla}_{\lambda} g_{\mu \nu}\,, \qquad
  Q_{\lambda}{}^{\mu \nu} = \bar{\nabla}_{\lambda} g^{\mu \nu} \,,
\end{align}
with symmetry brackets defined as $S_{[ab]} = (S_{ab}-S_{ba})/2$ and $S_{(ab)} = (S_{ab}+S_{ba})/2$.
From the metrical structure, we also have the Christoffel symbols associated with a Levi-Civita connection
\begin{equation}
\Gamma^{\lambda}_{\mu \nu} = \frac{1}{2} g^{\lambda \rho} \left( \partial_{\mu} g_{\nu \rho} + \partial_{\nu} g_{\mu \rho} - \partial_{\rho} g_{\mu \nu} \right) \, ,
\end{equation}
which is metric-compatible $\nabla_{\mu}g_{\nu \lambda} =0$ and torsion-free $\Gamma^{\lambda}_{[\mu \nu]} = 0$.
The affine connection can be decomposed into the sum of the Levi-Civita connection and the contortion tensor~\cite{JS1954,Hehl:1994ue}
\begin{align}
  \label{contorsion}
  \bar{\Gamma}{}^{\lambda}_{\mu \nu} =
  \Gamma{}^{\lambda}_{\mu \nu} + K_{\mu \nu}{}^{\lambda} \,, \qquad
  K_{\mu \nu}{}^{\lambda} = \frac{1}{2}  g^{\lambda \rho}
  \bigl( -T_{\{\nu \mu \rho \}} + Q_{\{ \mu \rho \nu \} } \bigr) \,,
\end{align}
where the \textit{Schouten bracket} permutes indices according to $
S_{\{ \mu \nu \lambda \} } = S_{\mu \nu \lambda} - S_{\nu \lambda \mu} + S_{\lambda \mu \nu}$.

From this decomposition, the affine Riemann tensor can be expressed in terms of the Levi-Civita curvature and additional contortion terms
\begin{align} \label{Riemann_K}
  \bar{R}_{\mu \nu \lambda}{}^{\rho} = R_{\mu \nu \lambda}{}^{\rho} +
  2 K_{[\mu| \sigma}{}^{\rho} K_{\nu] \lambda}{}^{\sigma} +
  2 \nabla_{[\mu} K_{\nu] \lambda}{}^{\rho} \, .
\end{align}
Contracting over appropriate indices gives the affine Ricci tensor $\bar{R}_{\mu \nu} = \bar{R}_{\lambda \mu \nu}{}^{\lambda}$ and further contracting with the metric gives the affine Ricci scalar $\bar{R} = g^{\mu \nu} \bar{R}_{\mu \nu}$. Applying this to the post-Riemannian expansion~(\ref{Riemann_K}) leads to
\begin{align} \label{Ricci_decomp}
\bar{R}_{\nu \gamma} &=  R_{\nu \gamma}+ 2 K_{[\lambda|\sigma}{}^{\lambda} K_{\nu] \gamma}{}^{\sigma} + 2\nabla_{[\lambda} K_{\nu] \gamma}{}^{\lambda} \, , 
\\
\bar{R} &=  R +  K_{\lambda \sigma}{}^{\lambda} K_{\nu}{}^{\nu \sigma} - K_{\nu \sigma}{}^{\lambda} K_{\lambda}{}^{\nu \sigma}  + \nabla_{ \lambda} K_{\nu}{}^{\nu \lambda} - \nabla_{\nu} K_{\lambda}{}^{\nu \lambda} 
\, .  \label{R_decomp}
\end{align}

A teleparallel geometry has vanishing total curvature $\bar{R}_{\gamma \mu \nu}{}^{\lambda} =0$ but non-vanishing torsion and non-metricity. Because the Riemann tensor is zero~(\ref{Riemann}), the affine connection is integrable and can be expressed as
\begin{equation} \label{Teleparallel_sol1}
\bar{\Gamma}^{\lambda}_{\mu \nu} = \Omega^{\lambda}{}_{\rho} \partial_{\mu} (\Omega^{-1})^{\rho}{}_{\nu} \, ,
\end{equation}
where $\Omega^{\rho}{}_{\mu}(x) \in GL(4,\mathbb{R})$, see for instance~\cite{BeltranJimenez:2018vdo,BeltranJimenez:2019esp}. The Ricci tensor and scalar also vanish, so that the Levi-Civita Ricci scalar can be written completely in terms of torsion and non-metricity
\begin{align}
0 =  R + K_{\lambda \sigma}{}^{\lambda} K_{\nu}{}^{\nu \sigma} - K_{\nu \sigma}{}^{\lambda} K_{\lambda}{}^{\nu \sigma}  + \nabla_{ \lambda} K_{\nu}{}^{\nu \lambda} - \nabla_{\nu} K_{\lambda}{}^{\nu \lambda} 
\, .  \label{R_tele}
\end{align}
The final two terms are total derivatives and so will not affect equations of motion, leading to an equivalence between the Einstein-Hilbert Lagrangian $R$ and an action consisting of the scalar contractions of contortion $ K_{\lambda \sigma}{}^{\lambda} K_{\nu}{}^{\nu \sigma} - K_{\nu \sigma}{}^{\lambda} K_{\lambda}{}^{\nu \sigma}$, as shown below.

Let us define the following scalar contractions of torsion and non-metricity, referred to as the torsion scalar, non-metricity scalar, and cross-terms respectively,
\begin{align} \label{T_scalar}
T &:= \frac{1}{4} T^{\mu \nu \lambda} T_{\mu \nu \lambda} + \frac{1}{2} T^{\mu \nu \lambda} T_{\nu \mu \lambda} - T^{\nu}{}_{\nu}{}^{\mu} T^{\lambda}{}_{\lambda \mu} \\
Q &:= \frac{1}{4} Q_{\mu \nu \lambda} Q^{\mu \nu \lambda} - \frac{1}{2} Q_{\mu \nu \lambda} Q^{\nu \lambda \mu} - \frac{1}{4} Q_{\mu}{}^{\nu}{}_{\nu} Q^{\mu \lambda}{}_{\lambda} + \frac{1}{2} Q_{\mu}{}^{\nu}{}_{\nu} Q_{\lambda}{}^{\mu \lambda} \, ,  \label{Q_scalar} \\
C &:= Q^{\mu \nu \lambda} T_{\nu \lambda \mu} + Q^{\mu \nu}{}_{\nu} T^{\lambda}{}_{\mu \lambda}
- Q^{\mu}{}_{\mu}{}^{\nu} T^{\lambda}{}_{\nu \lambda}  \, .\label{C_scalar}
\end{align}
By also introducing the torsion and non-metricity `superpotentials' 
\begin{align} 
 \mathscr{S}^{\mu \nu \lambda} &:=  \frac{1}{2} \frac{\partial T}{\partial T_{\mu \nu \lambda}}  =  \frac{1}{4} \big( T^{\mu \nu \lambda} + T^{\nu \mu \lambda} + T^{\lambda \nu \mu}  \big) -  \frac{1}{2} \big( g^{\mu \nu} T^{\rho}{}_{\rho}{}^{\lambda} -g^{\mu \lambda} T^{\rho}{}_{\rho}{}^{\nu} \big) \, ,\label{T_superpotential} \\
\mathscr{P}^{\lambda \rho \sigma}  &:= \frac{1}{2} \frac{\partial Q}{\partial Q_{\lambda \rho \sigma}} = \frac{1}{4}Q^{\lambda \rho \sigma} - \frac{1}{2} Q^{(\rho \sigma)\lambda} - \frac{1}{4} g^{\rho \sigma}( Q^{\lambda \mu}{}_{\mu} - Q^{\mu}{}_{\mu}{}^{\lambda}) + \frac{1}{4} g^{\lambda (\rho} Q^{\sigma) \mu}{}_{\mu} \, , \label{Q_superpotential} 
\end{align}
the scalar equations~(\ref{T_scalar})--(\ref{C_scalar}) can be written as $T =  \mathscr{S}^{\mu \nu \lambda} T_{\mu \nu \lambda}$, $Q= \mathscr{P}^{\mu \nu \lambda} Q_{\mu \nu \lambda}$ and $C = 2  \mathscr{S}^{\mu \nu \lambda} Q_{\lambda \nu \mu}$. One can then check that these scalar contractions are simply the sum of the quadratic contortion tensor terms
\begin{equation}
 K_{\lambda \sigma}{}^{\lambda} K_{\nu}{}^{\nu \sigma} - K_{\nu \sigma}{}^{\lambda} K_{\lambda}{}^{\nu \sigma} = T + Q +  C \, .
\end{equation}
Finally, we introduce the torsion and non-metricity boundary terms 
\begin{align}
B_T &:= \frac{2}{\sqrt{-g}} \partial_{\mu} \left(\sqrt{-g} T^{\lambda \mu}{}_{\lambda}\right) \, , \label{BT} \\
B_Q &:= \frac{2}{\sqrt{-g}} \partial_{\mu} \left( \sqrt{-g} Q^{[\lambda \mu]}{}_{\lambda} \right) \, .  \label{BQ}
\end{align}
from which we have $2\nabla_{\mu} K_{\lambda}{}^{[\lambda \mu]} = B_T + B_Q$.
With these definitions, we can rewrite the teleparallel equation~(\ref{R_tele}) relating the Levi-Civita Ricci scalar to the geometric scalars and boundary terms
\begin{equation} \label{fundamental}
0 = R + T + Q + C + B_T + B_Q \, .
\end{equation}
Noting that boundary terms will not contribute to variations within an action, we state the teleparallel equivalent of General Relativity (TEGR) as the torsional and non-metric parts of~(\ref{fundamental}) without the boundary terms  
\begin{equation}
S_{\rm{TEGR}} = -\frac{1}{2 \kappa} \int \left(T + Q +C \right) \sqrt{-g} d^4x + \int \bar{R}_{\mu \nu \lambda}{}^{\rho} r^{\mu \nu \lambda}{}_{\rho} d^4x  \, .
\end{equation}
The final term is a Lagrange multiplier enforcing the teleparallel constraint~(\ref{Teleparallel_sol1}), see~\cite{Adak:2008gd,BeltranJimenez:2018vdo} for details.
Here we also assume that the manifold in question is compact, such that the equivalence with GR does not require additional boundary terms or conditions. For other considerations, see~\cite{Oshita:2017nhn,Koivisto:2022oyt}. Throughout this work, we continue to assume the manifold is compact.
 
In the case that only torsion contributes, with non-metricity set to zero, this is referred to as metric teleparallel gravity. Alternatively, one can set torsion to zero and focus only on non-metricity, which is known as symmetric teleparallel gravity. This is the geometric framework we will be most interested in throughout this work, and the one in which we will construct modifications to GR where the equivalence in~(\ref{fundamental}) is broken.

\subsection{Symmetric teleparallel gravity}

In the symmetric teleparallel framework with vanishing torsion $T^{\lambda}{}_{\mu \nu} =0$, Eq.~(\ref{fundamental}) reduces to
\begin{equation} \label{RQB}
R = -Q - B_{Q} \, .
\end{equation}
The symmetric teleparallel equivalent of General Relativity (STEGR) is then described by the action\footnote{Note that we have left out Lagrange multipliers from this action. From this point onwards, we will assume that the symmetric teleparallel connection is being used, as explained directly below.}
\begin{equation}
S_{\rm{STEGR}} = -\frac{1}{2 \kappa} \int Q \sqrt{-g} d^4 x \, ,
\end{equation}
with the resulting metric variations leading to the Einstein tensor $G_{\mu \nu} = R_{\mu \nu} - \frac{1}{2} g_{\mu \nu} R$. Variations with respect to the connection vanish, which can be seen as a consequence of~(\ref{RQB}): the Levi-Civita Ricci scalar depends only on the metric so its affine connection variations are trivially zero, while $B_Q$ is a boundary term so its variations must also vanish.
 To be more explicit, for this geometry the affine connection with vanishing curvature and torsion can be written of the form~\cite{BeltranJimenez:2022azb}
\begin{equation} \label{stg_connection}
\bar{\Gamma}^{\gamma}_{\mu \nu} = \frac{\partial x^{\gamma}}{\partial \xi^{\rho}}  \partial_{\mu} \partial_{\nu} \xi^{\rho} \, ,
\end{equation}
where the fields $\xi^{\mu}$ parametrise the connection\footnote{Note that $\xi^{\mu} = (\xi^1,\xi^2,\xi^3,\xi^4)$ is \textit{not} a vector, in the same way that coordinates $x^{\mu}$ are not vectors.}, and the first term should be understood as the inverse of the matrix $(\partial \xi^{\rho} / \partial x^{\gamma})$. For a specific choice of coordinates where $\xi^{\mu} = x^{\mu}$, the connection trivialises and~(\ref{stg_connection}) vanishes $\bar{\Gamma}^{\gamma}_{\mu \nu}=0$. This particular coordinate choice is known as the \textit{coincident gauge}, or unitary gauge~\cite{Adak:2008gd,BeltranJimenez:2017tkd}. A holonomic coordinate system where the coefficients of the affine connection vanishes at all points exists if and only if the connection is symmetric and has vanishing curvature tensor~\cite{eisenhart1927non}. The symmetric teleparallel geometry can therefore be defined by the existence of such a gauge while working in the coordinate basis, in the same way that the Weitzenb\"{o}ck connection, i.e., the vanishing spin connection in the orthonormal basis $\omega^{a}{}_{b}=0$, defines the metric teleparallel geometry~\cite{Jensko:2023lmn,Golovnev:2020zpv}. Performing a general coordinate transformation from the vanishing coincident gauge connection to an arbitrary coordinate system $\xi^{\mu} \rightarrow x^{\mu}$,
\begin{align}
 \bar{\Gamma}^{\gamma}_{\mu \nu}(\xi) \rightarrow \hat{\bar{\Gamma}}^{\gamma}_{\mu \nu}(x) &=  \frac{\partial \xi^{\alpha}}{\partial x^{\mu}} \frac{\partial \xi^{\beta}}{\partial x^{\nu}} \frac{\partial x^{\gamma}}{\partial \xi^{\sigma}} \bar{\Gamma}^{\sigma}_{\alpha \beta}(\xi) + \frac{\partial^2 \xi^{\sigma}}{\partial x^{\mu} \partial x^{\nu} } \frac{\partial x^{\gamma}}{\partial \xi^{\sigma}} \nonumber \\
 &=\frac{\partial x^{\gamma}}{\partial \xi^{\sigma}} \partial_{\mu} \partial_{\nu} \xi^{\sigma} \, ,
\end{align}
leads precisely to the form given in~(\ref{stg_connection}). Transforming instead from an arbitrary coordinate system to the coincident gauge $x^{\mu} \rightarrow \hat{x}^{\mu} = \xi^{\mu}$, we simply perform the reverse transformation 
\begin{align}
	\bar{\Gamma}^{\gamma}_{\mu \nu}(x) \rightarrow \hat{\bar{\Gamma}}^{\gamma}_{\mu \nu}(\xi) &=  \frac{\partial x^{\alpha}}{\partial \xi^{\mu}} \frac{\partial x^{\beta}}{\partial \xi^{\nu}} \frac{\partial \xi^{\gamma}}{\partial x^{\sigma}} \bar{\Gamma}^{\sigma}_{\alpha \beta}(x) -  \frac{ \partial^2 \xi^{\gamma}}{\partial x^{\alpha} x^{\beta}} \frac{  \partial x^{ \alpha}}{ \partial \xi^{\mu}} \frac{ \partial x^{\beta}}{ \partial \xi^{\nu}}  
	 \nonumber \\
	&=   \frac{\partial x^{\alpha}}{\partial \xi^{\mu}} \frac{\partial x^{\beta}}{\partial \xi^{\nu}}  \left( \frac{\partial \xi^{\gamma}}{\partial x^{\sigma}} \bar{\Gamma}^{\sigma}_{\alpha \beta} - \partial_{\alpha} \partial_{\beta} \xi^{\gamma} \right) = 0 \, ,
\end{align}
with the final line vanishing by the definition~(\ref{stg_connection}).

 In the coincident gauge, covariant derivatives reduce to partial derivatives, and the non-metricity tensor is simply the partial derivative of the metric
\begin{equation}
\bar{\Gamma}^{\lambda}_{\mu \nu} \rightarrow 0 \, , \qquad \bar{\nabla}_{\mu}  \rightarrow  \partial_{\mu}   \,, \qquad Q_{\mu \nu \lambda} \rightarrow \mathring{Q}_{\mu \nu \lambda} = - \partial_{\mu} g_{\nu \lambda} \, .
\end{equation}
We will denote geometric objects with an over-ring to signal that they are in the coincident gauge. In following section we explore this gauge in more detail. For now, we emphasize that this gauge is \textit{always} available, given some appropriate coordinate transformation. The vanishing affine connection may seem to imply that some dynamics have been lost, or that the total degrees of freedom must necessarily be reduced, but that is not the case, as we aim to show. In the STEGR limit, this has no impact --- the connection itself being non-dynamical --- but it becomes important for non-linear modifications beyond GR.
This gauge (coordinate) choice does not change to the underlying physics, with all physical predictions and observables being unaffected. However, it reveals fundamental information about the mathematical structure of these geometries.

\subsubsection*{$f(Q)$ gravity}
The modified action of symmetric teleparallel gravity is taken to be
\begin{equation} \label{Q_action}
S = -\frac{1}{2\kappa} \int f(Q) \sqrt{-g} d^4x \, ,
\end{equation}
where again curvature and torsion have been set to zero $\bar{R}_{\mu \nu \gamma}{}^{\lambda} = T^{\lambda}{}_{\mu \nu}=0$. Note that these constraints can be implemented in various different ways, such as via Lagrange multipliers, restricted variations, or solving explicitly for the teleparallel connection (such as in Eq.~(\ref{stg_connection})). If implemented properly, all of these routes are equivalent, and we refer to~\cite{Gomes:2023hyk} for more explicit details. The key point to remember is that for the non-linear modifications, the metric and the affine connection (or fields derived from it~\cite{Nojiri:2024zab,Nojiri:2024hau}) must be varied independently, as these represent additional degrees of freedom beyond the metric. This is a simple consequence of the relation between the Levi-Civita Ricci scalar and the non-metricity scalar (\ref{RQB}), which will only be equivalent (modulo boundary terms) in the linear theory. The explicit dependence of the action on the connection will be shown below.

From the metric variations we obtain the following field equations
\begin{equation}
	f'(Q) \big( G_{\rho \sigma} - \frac{1}{2} g_{\rho \sigma} Q \big) + 2 f''(Q) \mathscr{P}^{\lambda}{}_{\rho \sigma} \partial_{\lambda} Q + \frac{1}{2} g_{\rho \sigma} f(Q) = \kappa T_{\rho \sigma} \, ,
\end{equation}
where $T_{\mu \nu}$ is the metric energy-momentum tensor.
From the connection variations one can derive the following equations~\cite{BeltranJimenez:2019tme}
\begin{equation} \label{f(Q)_connection}
	\bar{\nabla}_{\mu} \bar{\nabla}_{\nu} \big( \sqrt{-g} \mathscr{P}^{\mu \nu}{}_{\lambda} f'(Q) \big) = \kappa  \bar{\nabla}_{\mu} \bar{\nabla}_{\nu} \left( \sqrt{-g} \Delta^{\mu \nu}{}_{\lambda} \right) \, ,
\end{equation}
with the hypermomentum defined as the connection variations of the matter action
\begin{equation}
\Delta^{\mu \nu}{}_{\lambda}  := -\frac{1}{2\sqrt{-g}}\frac{\delta S_{\textrm{m}}}{\delta \bar{\Gamma}^{\lambda}_{\mu \nu}} \, .
\end{equation}
These connection field equations are proportional to the Levi-Civita covariant derivative of the metric field equations, which can be derived straightforwardly by looking at infinitesimal diffeomorphisms of the action~\cite{BeltranJimenez:2019tme}.
For vanishing hypermomentum, the energy-momentum tensor is covariantly conserved $\nabla_{\mu} T^{\mu \nu} = 0$. It is usually assumed that matter does not couple to the connection (or couples in such a way that the right-hand side of~(\ref{f(Q)_connection}) vanishes, which has been shown to be true for standard matter~\cite{BeltranJimenez:2020sih}). However, we refer to~\cite{Iosifidis:2020gth,Hohmann:2021ast} for studies considering non-trivial hypermomentum. In this paper, we continue to assume that matter couples only to the Levi-Civita connection; this could be considered somewhat unnatural in a true metric-affine formulation, but these issues become redundant in the gauge-fixed formalism below.

\subsection{Gauge-fixed formalism} \label{sec:2.2}
Let us briefly introduce a gauge-fixed formalism for the teleparallel theories. This approach was taken in our previous work~\cite{Boehmer:2021aji}, and generalised to a metric-affine geometry in~\cite{Boehmer:2023fyl}. In the former case, the equivalence with the teleparallel theories should be clear~\cite{BeltranJimenez:2022azb}. The latter case leads instead to a generalised metric-affine type of theory, which has neither GR nor teleparallel gravity as a limit. Here, we will use this gauge-fixed formalism to work in the coincident gauge for the $f(Q)$ theories, making the equivalence transparent.

Another way to interpret the explanation behind this gauge is by formulating teleparallel gravity as a Riemannian theory (in the standard Levi-Civita geometry), where the Ricci scalar has been decomposed into a bulk and boundary part
\begin{equation}
R = \ourG + \ourB \, .
\end{equation}
 This decomposition is basis dependent, and the resulting bulk and boundary terms are non-covariant. Working in the coordinate basis\footnote{This should be stated explicitly, as working in different bases leads to a different type of action~\cite{Hehl:1979gp}.}, this bulk term gives rise to the so-called \textit{Einstein action} or \textit{Gamma-squared action}
 \begin{equation} \label{E_action}
S_{E} = \frac{1}{2 \kappa} \int \ourG \sqrt{-g} d^4x  =  \frac{1}{2 \kappa} \int  g^{\mu \nu} \left( \Gamma^{\lambda}_{\mu \rho} \Gamma^{\rho}_{\nu \lambda} -  \Gamma^{\lambda}_{\lambda \rho} \Gamma^{\rho}_{\mu \nu} \right) \sqrt{-g} d^4x \, ,
\end{equation}
which is first-order and non-covariant. It is well-known that this Lagrangian is exactly the non-metricity scalar $Q$ in the coincident gauge~\cite{BeltranJimenez:2019esp}. Moreover, the bulk term differs from the full non-metricity scalar by yet another boundary term~\cite{Boehmer:2021aji}
 \begin{equation} \label{G_decomp}
 \ourG(g) \equiv - \mathring{Q}(g) = - Q(g,\bar{\Gamma}) - b_{Q}(g,\bar{\Gamma}) \, , \qquad  b_{Q}(g,\bar{\Gamma})  := \frac{2}{\sqrt{-g}} \partial_{\mu} \left(\sqrt{-g} g^{\lambda [\nu} \bar{\Gamma}^{\mu]}_{\nu \lambda} \right) \, ,
 \end{equation}
 where we have defined $\mathring{Q}$ as the coincident gauge-fixed non-metricity scalar, and also made explicit the dependence of these objects on the metric and affine connection.
In fact, without making reference to teleparallel gravity, it can be shown that the non-covariant Einstein action can be made covariant via the Stueckelberg trick~\cite{BeltranJimenez:2022azb}. First, performing a general coordinate transformation gives
\begin{align} \label{G_stueck}
\hat{\ourG} =  \ourG - E_{\gamma}{}^{\beta \kappa} \Big( \frac{\partial x^{\gamma}}{ \partial \hat{x}^{\mu}} \frac{\partial^2 \hat{x}^{\mu}}{\partial x^{\beta} \partial x^{\kappa} }
  \Big)  
 + g^{\alpha \beta} \big(\delta^{\sigma}_{\beta} \delta^{\eta}_{\gamma} - \delta^{\eta}_{\beta} \delta^{\sigma}_{\gamma} \big) \frac{\partial x^{\kappa}}{\partial \hat{x}^{\mu}} \frac{\partial^2 \hat{x}^{\mu}}{\partial x^{\eta} \partial x^{\alpha} } \frac{\partial x^{\gamma}}{\partial \hat{x}^{\rho}}   \frac{\partial^2 \hat{x}^{\rho}}{\partial x^{\sigma} \partial x^{\kappa} }  \, ,
\end{align}
where $E_{\gamma}{}^{\beta \kappa}$ is a linear combination of Levi-Civita Christoffel symbols, defined in~(\ref{E}) below.
Promoting the new coordinates to Stueckelberg fields $\hat{x}^{\mu} \rightarrow \xi^{\mu}$ and comparing with the symmetric teleparallel connection~(\ref{stg_connection}) shows us that the covariantisation procedure gives exactly the non-metricity scalar,
\begin{equation}
\hat{\ourG} =  \ourG + b_{Q} = -Q  \, ,
\end{equation}
where the terms on the right of~(\ref{G_stueck}) have been identified with $b_Q$ in Eq.~(\ref{G_decomp}),
see Appendix D of~\cite{Jensko:2023lmn} for technical details.

Taking the non-covariant Einstein action and studying the theory
\begin{equation} \label{G_action}
\int f(\ourG) \sqrt{-g} d^4 x \, ,
\end{equation}
is the approach taken in~\cite{Boehmer:2021aji}, and this leads precisely to $f(Q)$ gravity in the coincident gauge. The metric field equations are\footnote{A redefinition $f(G) \rightarrow - f(Q)$ is necessary to account for the sign difference between the actions~(\ref{Q_action}) and~(\ref{G_action}).}
\begin{align}
\label{metric_EoM}
	f'(\ourG) \left( G_{\rho \sigma} + \frac{1}{2} g_{\rho \sigma} \ourG \right) + \frac{1}{2} f''(\ourG) E_{\rho \sigma}{}^{\gamma} \partial_{\gamma} \ourG - \frac{1}{2} g_{\rho \sigma} f(\ourG) =
	\kappa T_{\rho \sigma} \,,
\end{align}
where the object $E^{\rho \sigma \gamma}$ is the non-covariant superpotential defined by
\begin{equation} \label{E}
	E^{\mu \nu \lambda} := 2 \frac{\partial \ourG}{\partial (g_{\mu \nu,\lambda})} =  2 g^{\rho \mu} g^{\nu \sigma} \Gamma^{\lambda}_{\rho \sigma} -
	2 g^{\lambda (\mu} g^{\nu) \sigma} \Gamma^{\rho}_{\rho \sigma} + g^{\mu \nu} g^{\lambda \rho} 
	\Gamma^{\sigma}_{\sigma \rho} - g^{\mu \nu} g^{\rho \sigma} \Gamma^{\lambda}_{\rho \sigma} \, .
\end{equation}
This object is symmetric over its first two indices $E^{[\mu \nu] \lambda} =0$ and from it we can reconstruct the bulk term $\ourG = \frac{1}{4}E^{\mu \nu \lambda} \partial_{\lambda} g_{\mu \nu}$. This term is simply the non-metricity superpotential~(\ref{Q_superpotential}) fixed to the coincident gauge
\begin{equation}
	\mathscr{P}_{\lambda \rho \sigma} \rightarrow \mathring{\mathscr{P}}_{\lambda \rho \sigma} = \frac{1}{4} E_{\rho \sigma \lambda} \, .
\end{equation}
For further properties and relations between these objects, we refer to~\cite{Boehmer:2021aji,Jensko:2023lmn}, but they will not be needed here.

The equation of motion for the connection in the covariant theory~(\ref{f(Q)_connection}) arises as a consistency condition relating to diffeomorphism invariance~\cite{BeltranJimenez:2019tme}. This can be seen plainly by making an infinitesimal coordinate transformation on the bulk term, which introduces the superpotential~\cite{Jensko:2023lmn}
\begin{equation} \label{Ginf}
\ourG \rightarrow \hat{\ourG} = \ourG + E_{\lambda}{}^{\mu \nu}  \partial_{\mu} \partial_{\nu} \xi^{\lambda} \, .
\end{equation}
Requiring the gravitational action to be diffeomorphism invariant then leads to
\begin{equation} \label{Econs}
\partial_{\mu} \partial_{\nu} \left( \sqrt{-g} E_{\lambda}{}^{\mu \nu} f'(\ourG) \right) = 0 \, ,
\end{equation}
which is simply~(\ref{f(Q)_connection}) in the coincident gauge
\begin{equation} \label{coincident_cons}
  \partial_{\mu} \partial_{\nu} \left( \sqrt{-g} \mathring{\mathscr{P}}^{\mu \nu}{}_{\lambda} f'(\mathring{Q}) \right) = 0 \, .
\end{equation}
From the perspective of the $f(\ourG)$ theories, the geometry is firmly Levi-Civita and their is no question about which connection matter should couple to. It is therefore natural for the right-hand side of~(\ref{Econs}), and hence~(\ref{coincident_cons}), to vanish. This is the case for all standard matter actions that are coordinate scalars, but see~\cite{Boehmer:2021aji} for further discussion. These conservation equations can be rewritten as
\begin{equation}
\nabla_{\mu} H^{\mu}{}_{\nu} = 0 \, ,
\end{equation}
where $H_{\mu \nu}$ represents the metric field equations. This then directly implies the covariant conservation of the energy-momentum tensor $\nabla_{\mu}T^{\mu}{}_{\nu} =0$. We will refer to this equation as the diffeomorphism constraint, or conservation equation, for obvious reasons.
This approach makes it possible to study the modified teleparallel theories without explicitly referencing geometries beyond standard GR. The inclusion of a boundary term in the modification $f(\ourG,\ourB)$ is also straightforward, as first studied in~\cite{Boehmer:2021aji}.

The gauge-fixed formulation is equally applicable to the $f(T)$ theories, but the decomposition of the Ricci scalar is instead computed in the orthonormal basis. Again, requiring the action to be locally Lorentz invariant gives rise to the antisymmetric equation corresponding to the spin-connection variations. This is well known from the pure versus the covariant approaches to $f(T)$ gravity~\cite{Krssak:2015oua}. For further explanation on these topics, and studies looking at the tetradic decomposition of the Ricci scalar which breaks local Lorentz invariance, see~\cite{Jensko:2023lmn} and references therein.

It should now be clear that the gauge-fixed and covariant theories are completely equivalent, and that solutions can always be mapped to the coincident (or Weitzenb\"{o}ck) gauge via appropriate coordinate (or local Lorentz) transformations. Any free functions and degrees of freedom of the teleparallel connections will instead manifest in the metric (or tetrad), resulting from the coordinate (or local Lorentz) transformations. This also shows that these theories have additional constraints to satisfy as a result of breaking fundamental symmetries, which would vanish identically for fully invariant theories such as GR or $f(R)$ gravity.

\section{Spatially curved FLRW cosmology}  \label{sec:kvf}
In this section we derive the coincident gauge coordinates for the FLRW spacetimes with and without spatial curvature. This allows us to state the form of the metric in these coordinates, thereby eliminating the need to consider the teleparallel connection (which trivialises for these coordinates only). Those wishing to skip directly to the $f(Q)$ equations of motion for these branches can look directly at Section~\ref{sec:fe}, where the coincident gauge metrics are also restated.

\subsection{Symmetries}
The Killing vectors of the homogeneous and isotropic cosmological symmetry in spherical coordinates $(t,r,\theta,\phi)$ are the rotation generators 
\begin{subequations}
\begin{align} \label{KV1}
\rho_{x} &= \sin{\phi} \, \partial_{\theta} + \frac{\cos \phi }{ \tan \theta } \partial_{\phi} \, , \\ \label{KV2}
\rho_{y} & = - \cos \phi \, \partial_{\theta} + \frac{ \sin{\phi}}{ \tan \theta} \partial_{\phi} \, , \\ \label{KV3}
\rho_{z} &= - \partial_{\phi} \, ,
\end{align}
\end{subequations}
and translation generators
\begin{subequations}
\begin{align} \label{KV4}
\tau_{x} & = \chi \sin{\theta} \cos{\phi} \,  \partial_{r} + \frac{\chi}{r} \cos{\theta} \cos{\phi} \, \partial_{\theta} -  \frac{\chi}{r}  \frac{\sin{\phi}}{\sin{\theta}} \partial_{\phi} \, , \\ \label{KV5}
\tau_{y} &=  \chi \sin{\theta} \sin{\phi} \,  \partial_{r} + \frac{\chi}{r} \cos{\theta} \sin{\phi} \, \partial_{\theta} +  \frac{\chi}{r}  \frac{\cos{\phi}}{\sin{\theta}} \partial_{\phi} \, , \\ \label{KV6}
\tau_{z} &= \chi \cos{\theta}  \,  \partial_{r} - \frac{\chi}{r} \sin{\theta} \, \partial_{\theta} \, ,
\end{align}
\end{subequations}
where $\chi := \sqrt{1 - k r ^2}$ and $k$ represents the spatial curvature~\cite{chrusciel2020elements}. We will collectively refer to the set of Killing vectors describing a given spacetime isometry by $Z_{\xi} = \left(\xi_{(1)}, ...,  \xi_{(n)} \right)$, where $\xi_{(i)}$ are the $n$ independent Killing vector fields\footnote{The Killing vectors $\xi_{(i)}$ should not be confused with the symbol used to represent the coordinate transformations of the previous section and the symmetric teleparallel connection.}. In the FLRW setting, we take the $\xi_{(i)}$ to be the rotation and translation generators given above.

Spacetime isometries of the metric satisfy the Killing equation~\cite{Hawking:1973uf}
\begin{equation} \label{L_met}
\mathcal{L}_{\xi} g_{\mu \nu} = \xi^{\lambda} \partial_{\lambda} g_{\mu \nu} + \partial_{\mu} \xi^{\lambda} g_{\lambda \nu} + \partial_{\nu} \xi^{\lambda} g_{\lambda \mu} =  2 \nabla_{(\mu} \xi_{\nu)} = 0 \, ,
\end{equation}
for each $\xi \in Z_{\xi}$. This is simply the vanishing Lie derivative of the metric tensor. In a metric-affine geometry, symmetries for the affine connection must satisfy~\cite{JS1954,Yano:2020}
\begin{align} \label{L_gamma}
\mathcal{L}_{\xi} \bar{\Gamma}^{\lambda}_{\mu \nu} &=  
\xi^{\rho} \partial_{\rho}  \bar{\Gamma}^{\lambda}_{\mu \nu} - \partial_{\rho} \xi^{\lambda}  \bar{\Gamma}^{\rho}_{\mu \nu} + \partial_{\mu} \xi^{\rho}  \bar{\Gamma}^{\lambda}_{\rho \nu} + \partial_{\nu} \xi^{\rho}  \bar{\Gamma}^{\lambda}_{\mu \rho} + \partial_{\mu} \partial_{\nu} \xi^{\lambda}  \nonumber \\
&= \bar{\nabla}_{\mu} \bar{\nabla}_{\nu} \xi^{\lambda} + \xi^{\rho} \bar{R}_{\rho \mu \nu}{}^{\lambda} - \bar{\nabla}_{\mu}(T^{\lambda}{}_{\nu \rho} \xi^{\rho}) \, ,
\end{align}
for all $\xi \in Z_{\xi}$. Our notion of spacetime symmetries follows that of~\cite{Hohmann:2019fvf}, where the Lie derivatives of both the metric and affine connection must vanish with respect to the given Killing vectors (sometimes called `generalised Killing vectors'~\cite{Obukhov:2015eha}). We will therefore not differentiate between metric isometries and isometries of the affine connection (which have also called `isoparallelisms'~\cite{Obukhov:2015eha} or `affine collineations'~\cite{Fonseca-Neto:1992xln}).

The metric that solves the Killing equation~(\ref{L_met}) for the homogenous and isotropic Killing vectors~(\ref{KV1})--(\ref{KV6}) is the FLRW metric, again stated in spherical coordinates, 
\begin{equation} \label{FLRW}
ds^2 = g_{\mu \nu} dx^{\mu} dx^{\nu}  = - N(t)^2 dt^2 + \frac{a(t)^2}{1-k r^2} dr^2 + a(t)^2 r^2 d \theta^2 + a(t)^2 r^2 \sin^2\theta d\phi^2 \, ,
\end{equation}
where $N(t)$ is the lapse and $a(t)$ is the scale factor~\cite{chrusciel2020elements}.
Analogously, one can also determine the non-vanishing components of an affine connection respecting these symmetries by solving~(\ref{L_gamma}). In the symmetric teleparallel case, the calculation is simplified due to the final two terms of~(\ref{L_gamma}) vanishing, see for instance~\cite{Hohmann:2019fvf}. We will take a slightly different route that does not explicitly refer to the components of the affine connection nor any geometric quantities beyond the metric. However, we will show the end result to be completely equivalent, as outlined below. 

Working in the symmetric teleparallel framework, the coincident gauge is defined as coordinates where the spacetime affine connection vanishes $\bar{\Gamma}^{\lambda}_{\mu \nu}=0$~\cite{eisenhart1927non}. This reduces the equation for the Lie derivative of the connection to
\begin{equation} \label{condKV}
\partial_{\mu} \partial_{\nu} \xi^{\lambda} = 0 \, .
\end{equation}
In other words, $\xi^{\lambda}(x)$ is an affine function of the coincident gauge coordinates. As such, this gauge condition is coordinate-dependent but still remains invariant under global Poincar\'{e} transformations. Eq.~(\ref{condKV}) can be seen as the defining condition of symmetries in the coincident gauge itself --- this should not be surprising when comparing with the form of the symmetric teleparallel connection~(\ref{stg_connection}). We now have an explicit way of calculating and checking that a coordinate system satisfies this constraint with respect to some given spacetime isometries.
Hence, any studies performed in the symmetric teleparallel theories such as $f(Q)$ gravity that claim the connection can be set to zero but where the Killing vectors are not affine functions of the chosen coordinate system must be invalid.

In an arbitrary, non-coincident gauge coordinate system, we instead have $\partial_{\mu} \partial_{\nu} \xi^{\lambda}  \neq 0$. Let us assume that our starting coordinate system is not in the coincident gauge. Performing a coordinate transformation and requiring the new system $\hat{x}^{\mu}$ to be the coincident gauge implies that 
\begin{equation} \label{KVcoinc}
\partial_{\mu} \partial_{\nu} \xi^{\lambda}  \rightarrow \hat{\partial}_{\mu}  \hat{\partial}_{\nu} \hat{\xi}^{\lambda} = 0 \, ,
\end{equation}
and this will be the set of equations for us to solve.
 In general, one may wonder whether~(\ref{KVcoinc}) is always solvable. In the case of the symmetric teleparallel geometry considered here, the answer is positive, as a result of forcing the affine connection to respect the isometries of spacetime~(\ref{L_gamma}). One can make the laborious calculation of fully expanding the general coordinate transformation~(\ref{KVcoinc}), which can then be rewritten in terms of the symmetric teleparallel connection~(\ref{stg_connection}). This once again results in the Lie derivative of the affine connection~(\ref{L_gamma}), which we demand to vanish. This is therefore a specific property of the symmetric teleparallel geometry, which relates to the integrability and symmetry of the affine connection~\cite{eisenhart1927non}. The form of Eq.~(\ref{KVcoinc}) is therefore somewhat deceiving because the gauge-fixing procedure obscures the geometric origin of this constraint.

\subsubsection{Gauge-fixed implementation}
The equivalent conditions can also be obtained without directly referencing the affine connection or its symmetries. This again makes use of the gauge-fixed formalism. To obtain~(\ref{condKV}) we instead begin with the Lie derivative of the bulk term $\ourG$ (which we remind the reader is simply the gauge-fixed non-metricity scalar $\ourG = - \mathring{Q}$), which takes the form
\begin{equation} \label{LieG}
\mathcal{L}_{\xi} \ourG = \xi^{\mu} \partial_{\mu} \ourG + E_{\lambda}{}^{\mu \nu} \partial_{\mu} \partial_{\nu} \xi^{\lambda} = 0 \, 
\end{equation}
for all $\xi \in Z_{\xi}$, and with $E_{\lambda}{}^{\mu \nu}$ defined in~(\ref{E}).
The final equality shows that the Lie derivative vanishes for isometries of the metric since $\mathcal{L}_{\xi} g = 0 \implies \mathcal{L}_{\xi} \Gamma = 0 \implies  \mathcal{L}_{\xi} \ourG = 0$. In other words, $\ourG$ is invariant under isometries of the metric because it depends only on the metric tensor and its derivatives~\cite{Yano:2020}. This is also not too difficult to check explicitly for the cosmological killing vectors~(\ref{KV1})--(\ref{KV6}).

We also know that the $f(\ourG)$ theories contain additional degrees of freedom, introduced by the explicit breaking of coordinate invariance. The Stueckelberg procedure outlined in Section~\ref{sec:2.2} makes this explicit, and the perturbative analysis of $f(\mathring{Q})$ gravity around cosmological backgrounds confirms the existence of at least two additional degrees of freedom\footnote{For discussion on the Hamiltonian approach we refer to~\cite{DAmbrosio:2023asf}.}~\cite{BeltranJimenez:2019tme}. However, in general these are hard to identify and somewhat hidden, as they are a consequence of diffeomorphism breaking. Nonetheless, all degrees of freedom (explicit or not) should be subject to the symmetries of spacetime. We can impose this on the hidden degrees of freedom by demanding their vanishing Lie derivative with respect to Killing vector fields. With covariance restored (i.e., making these degrees of freedom explicit), the object $\ourG$ should transform like a scalar. This is tantamount to the condition that
\begin{equation}
	\mathcal{L}_{\xi} \ourG = \xi^{\mu} \partial_{\mu} \ourG = 0 \, ,
\end{equation}
which implies that $E_{\lambda}{}^{\mu \nu} \partial_{\mu} \partial_{\nu} \xi^{\lambda} = 0$. The non-covariant object $E_{\lambda}{}^{\mu \nu}$ is a linear combination of Levi-Civita Christoffel symbols, which will not vanish non-locally in any coordinate system unless the spacetime is (Levi-Civita) flat $g_{\mu \nu} = \eta_{\mu \nu}$. Moreover, the vanishing of $E_{\lambda}{}^{\mu \nu}$ implies that $\ourG$ is also zero. Therefore, in all other cases we are led to $\partial_{\mu} \partial_{\nu} \xi^{\lambda} = 0$. This same argument applies to any non-covariant theories constructed from partial derivatives of the metric $\partial_{\mu} g_{\nu \lambda}$, or equally $\Gamma^{\lambda}_{\mu \nu}$, which leads to $\partial_{\mu} \partial_{\nu} \xi^{\lambda} = 0$. This follows directly from the transformation rules for the Christoffel symbols.

 Returning to the full gauge-invariant non-metricity scalar, spacetime isometries, (2.4) and (2.5), imply that 
 \begin{equation}
 \mathcal{L}_{{\xi}} Q =  {\xi}^{\mu} \partial_{\mu} Q = 0 \, ,
 \end{equation}
 which reduce to $ {\xi}^{\mu} \partial_{\mu} \mathring{Q} = -{\xi}^{\mu} \partial_{\mu} \ourG$ in the coincident gauge. Hence, applying the same reasoning as before, we see from~(\ref{LieG}) that in the coincident gauge one again has 
\begin{equation}
\partial_{\mu} \partial_{\nu} \xi^{\lambda} = 0\, ,
\end{equation}
 and the gauge-fixed formulation yields the same results.
 
  A quick calculation using the cosmological Killing vectors in spherical coordinates~(\ref{KV1})--(\ref{KV6}) indeed shows that these are incompatible with the coincident gauge. Taking $\tau_z$ as an example, we have
 \begin{equation}
\tau_{z}^{\mu} \partial_{\mu} \ourG = - \frac{4 \cos \theta}{r^3 a^2} \neq 0 \, ,
 \end{equation}
 and similarly for the other translation generators. In the next section we will find the appropriate transformation to the coincident gauge where all of these equations will vanish.

\subsection{Transformation to the coincident gauge}
In this section we derive the coincident gauge coordinates for the FLRW spacetimes with and without spatial curvature. By working in these unitary coordinates, we avoid complications associated with non-trivial connections. All dynamics and free functions will then reside in the metric tensor alone, allowing for a more straightforward comparison between the flat and curved branches.

The rotation and translation generators do not satisfy the condition $\partial_{\mu} \partial_{\nu} \xi^{\lambda} = 0$, except for $\rho_z$~(\ref{KV3}). As previously explained, the Killing vectors must take the form of affine functions of the coordinates, $a^{\mu}{}_{\nu} x^{\nu} + b^{\mu}$ where the components of $a^{\mu}{}_{\nu}$ and $b^{\mu}$ are constants. We are therefore forced to find coordinates `adapted' to the Killing vectors. One example is well-known in the literature: the spatially flat $k=0$ standard Cartesian coordinates. In these coordinates, the affine connection equations of motion are identically satisfied and there are no additional constraints to be solved. Another recently studied case is spherical symmetry~\cite{Bahamonde:2022zgj}, which expectedly takes a much more complicated form than in standard coordinates. 

One may also then be tempted to attribute some physical significance to these coincident gauge coordinates, which are adapted to the physical symmetries of the spacetime in a natural way. Here we refrain from further speculation on this matter, but note that this approach does give a definitive, systematic way of finding these coincident gauge coordinates. This could be utilised for studying energy-momentum pseudotensors, which are inherently coordinate-dependent. In fact, this is related to the original motivation for considering the symmetric teleparallel geometries~\cite{Nester:1998mp}. For further discussion on this topic, we refer to~\cite{BeltranJimenez:2019bnx,Gomes:2023hyk}.

Returning to the curved FLRW metric isometries and Killing vectors, we first aim to find a transformation such that the second partial derivative vanishes, see Eq.~(\ref{KVcoinc}). To do so, we begin with the following ansatz used by Hohmann~\cite{Hohmann:2021ast}
\begin{equation} \label{Anzatz}
\hat{x}^{0} = T(t,r) \, , \quad \hat{x}^{1} = R(t,r) \sin \theta \cos \phi \, , \quad \hat{x}^{2} = R(t,r) \sin \theta \sin \phi \, , \quad \hat{x}^{3} = R(t,r) \cos \theta \, ,
\end{equation}
where $T$ and $R$ are arbitrary functions of $t$ and $r$. The Jacobian transformation matrix is then given by
\begin{equation}
	\frac{\partial \hat{x}^{\mu}}{\partial x^{\nu}} = \begin{pmatrix}
	 	\partial_t T & \partial_r T & 0 & 0\\
		 \sin \theta \cos \phi \partial_t R  &  \sin \theta \cos \phi \partial_r R & \cos \theta \cos \phi R & - \sin \theta \sin \phi R \\
		 \sin \theta \sin \phi \partial_t R & \sin \theta \sin \phi \partial_r R & \cos \phi \sin \theta R & \sin \theta \cos \phi R \\
		 \cos \theta \partial_t R & \cos \theta \partial_r R & - \sin \theta R & 0 
	\end{pmatrix} \, .
\end{equation}
The coordinate transformation reduces to the standard polar-to-Cartesian transformation of flat space when $R(t,r) = r$ and  $T(t,r) = t$, and so the new coordinates for arbitrary $R$ and $T$ can be thought of as being related to the Cartesian one. In Hohmann's work~\cite{Hohmann:2021ast}, the functions $R$ and $T$ are determined in the coincident gauge by imposing the vanishing of the affine connection; the non-vanishing connection coefficients were calculated previously from the connection Killing equations with the cosmological symmetries~(\ref{L_gamma}), see also~\cite{Hohmann:2019fvf}. Here, we instead work directly with the equations $\partial_{\mu} \partial_{\nu} \xi^{\lambda} =0$, leading to a completely different set of equations to be solved and bypassing the affine connection and its symmetries altogether. Nonetheless, we verify that the final results are consistent with~\cite{Hohmann:2021ast}.

After performing the coordinate transformation above~(\ref{Anzatz}), the rotation Killing vectors have the following transformed components 
\begin{subequations}
\begin{align}
	\hat{\rho}_{x} &=(0,0, \cos \theta R, -\sin{\theta} \sin{\phi} R ) \, ,  \\
	\hat{\rho}_{y} &=(0,-\cos \theta R, 0, \sin \theta \cos \phi R ) \, , \\
	\hat{\rho}_{z} &= (0, \sin \theta \sin \phi R, - \sin \theta \cos \phi R, 0 )  \, , 
\end{align}
\end{subequations}
which can be written in terms of the new coordinate system $ \hat{x}^{\mu}(x) = Y^{\mu} = (\tilde{t},x,y,z)$ simply as
\begin{subequations}
\begin{align}
	\rho_{x}(Y) &=(0,0,z,-y) \, ,  \\
	\rho_{y}(Y) &=(0,-z,0,x ) \, , \\
	\rho_{z}(Y) &= (0,y,-x,0 )  \, .
\end{align}
\end{subequations}
This is the standard Cartesian form of the FLRW rotation generators, and these satisfy the coincident gauge constraints $\hat{\partial}_{\mu} \hat{\partial}_{\nu} \hat{\rho}^{\lambda}_{i} = 0$, where here the index $i$ counts the three independent rotations.

The translation Killing vector components take a more complicated form
\begin{subequations}
\begin{align}
	\hat{\tau}_{x} &=  \begin{pmatrix}
		\chi \sin \theta \cos \phi \partial_{r} T  \\
		  \frac{\chi R}{4r} \big(3+ \cos 2 \theta - 2 \sin^2 \theta \cos 2 \phi \big) + \chi \sin^2 \theta \cos^2 \phi \partial_{r} R  
		 \\
	 \frac{\chi}{r} \sin^2 \theta \sin \phi \cos \phi (r \partial_{r} R - R),   \\
	\frac{\chi}{r} \sin \theta \cos \theta \cos \phi (r \partial_{r} R - R)
	\end{pmatrix} \, , \\
		\hat{\tau}_{y} &=  \begin{pmatrix}
	 	 \chi \sin \theta \sin \phi \partial_{r} T \\
\frac{\chi}{r} \sin^2 \theta \sin \phi \cos \phi (r \partial_{r} R - R)
\\
\frac{\chi R}{4r} \big(3+ \cos 2 \theta + 2 \sin^2 \theta \cos 2 \phi \big) + \chi \sin^2 \theta \sin^2 \phi \partial_{r} R
\\ \frac{\chi}{r} \sin \theta \cos \theta \sin \phi (r \partial_{r} R - R) 
	\end{pmatrix} \, , \\
	\hat{\tau}_{z} &=  \begin{pmatrix}
 \chi \cos \theta  \partial_{r} T \\
\frac{\chi}{r} \sin \theta \cos \theta \cos \phi (r \partial_{r} R - R)\\  \frac{\chi}{r} \sin \theta \cos \theta \sin \phi (r \partial_{r} R - R)
\\
\frac{\chi}{r} \sin^2 \theta R + \chi \cos^2 \theta \partial_{r} R
	\end{pmatrix} \, .
\end{align}
\end{subequations}
Unlike the rotation Killing vectors, the terms with derivatives of the functions $T$ and $R$ immediately show that the translation vectors cannot be written as an affine function of the new coordinates. This instead leads to a set of differential equations for $R$ and $T$ that need to be solved. In total, there are 64 equations for each of the three translation generators
\begin{equation} \label{t_eqs}
	\sum_{i=1}^3 \hat{\partial}_{\mu} \hat{\partial}_{\nu} \hat{\tau}^{\lambda}_{i} = 0
\end{equation}
though these are not all independent. These sets of equations will be studied in the spatially flat and spatially curved cases individually, with further details given in Appendix \ref{Append0}. We begin with the spatially flat branches.

\subsubsection{Spatially flat $k=0$ solutions}

In Appendix \ref{Append01} we solve the differential equation~(\ref{t_eqs}) for the spatially flat $k=0$ case. This yields the partial solutions
\begin{equation} \label{partialSols1}
R(t,r) = r R(t) \, ,  \qquad T(t,r) = \frac{1}{2}r^2 T_1 (t) + T_2 (t)  \, ,
\end{equation}
and leaves two remaining sets of equations to be solved
	\begin{align}   \label{Tdiffeq1Main}
		T_1(t) \dot{R}(t) = R(t) \dot{T}_1(t) &= 0 \, , \\
		\dot{T}_2(t) \ddot{R}(t) - \dot{R}(t) \ddot{T}_2(t) &= 0 \, . \label{Tdiffeq2Main}
	\end{align}
These give two branches of solutions for either $T_1 = 0$,
or for $\dot{R}=0$ and $\dot{T}_1=0$.
We also include a limiting case of both branches separately, with $T_1 = \dot{R} = 0$, for reasons that will become clear. Lastly, there is the requirement that $T(t,r)$ must depend on $t$, and that $R(t)$ must be non-zero, such that the coordinate transformation is regular. This should be kept in mind throughout.

\begin{enumerate}
\item 
For the first solution with $T_1(t) = 0$, the first equations are satisfied~(\ref{Tdiffeq1Main}) and the second~(\ref{Tdiffeq2Main}) can be solved to give
\begin{equation} \label{T2sol}
T_2(t) = c_1 + c_2  R(t) \, .
\end{equation}
It is then straightforward to show explicitly that the coincident gauge conditions are satisfied $\hat{\partial}_{\mu} \hat{\partial}_{\nu} \hat{\tau}^{\lambda}_{i}=0$ for all the translation generators $\hat{\tau}_{i}$.
The  coincident gauge coordinate transformation for this branch and class of solutions is given by
\begin{equation} \label{coordset1}
	\hat{x}^{0} = R(t)  \, , \quad \hat{x}^{1} = r R(t)   \sin \theta \cos \phi \, , \quad \hat{x}^{2} = r R(t)  \sin \theta \sin \phi \, \qquad \hat{x}^{3} = r R(t)  \cos \theta \, ,
\end{equation}
where we have fixed the constants $c_1=0$ and $c_2=1$ without loss of generality\footnote{Note that we are always still free to make global affine coordinate transformations without affecting the structure of the coincident gauge.}. The components of the translation Killing vectors transform to
\begin{subequations}
\begin{align}
	\hat{\tau}_{x} &= \big(0, R(t), 0 ,0 \big) \, ,  \\
	\hat{\tau}_{y} &= \big(0,0,R(t),0 \big) \, , \\
	\hat{\tau}_{z} &= \big(0,0,0,R(t)\big) \, .
\end{align}
\end{subequations}
In the coincident gauge coordinates, which we label $\hat{x}^{\mu} = Y^{\mu}_{(1)} = (\tilde{t},x,y,z)$ for this branch, they simplify to
\begin{subequations}
\begin{align}
	\tau_{x}(Y) &= (0, \tilde{t} ,0,0 ) \, ,\\
	\tau_{y}(Y) &= (0,0,\tilde{t},0 ) \,,  \\
	\tau_{z}(Y) &= (0,0,0, \tilde{t} ) \, .
\end{align}
It is interesting that this transformation mixes space and time coordinates, with each translation vector now including a $\tilde{t}$ term. Contrast this with the usual Cartesian translation Killing vectors, where each $\tilde{t}$ term would be replaced with a constant.
\end{subequations}

\item
The second class of solutions is for $R(t) = c_3$, such that the second equation~(\ref{Tdiffeq2Main}) is satisfied, and then $T_{1}(t) = c_4$ satisfies the first equations~(\ref{Tdiffeq1Main}).
After which, one can again verify that $\hat{\partial}_{\mu} \hat{\partial}_{\nu} \hat{\tau}^{\lambda}_{i}=0$ for all Killing vectors. The coordinate transformation for this class of solution takes the form
\begin{equation}  \label{coordset2}
	\hat{x}^{0} = T_{2}(t) + \frac{r^2}{2}  \, , \quad \hat{x}^{1} = r \sin \theta \cos \phi \, , \quad \hat{x}^{2} = r \sin \theta \sin \phi \, , \quad \hat{x}^{3} = r \cos \theta \, ,
\end{equation}
where we have again fixed the constants $c_4=1 = c_3=1$. The translation Killing vectors have the following components
\begin{subequations}
\begin{align}
	\hat{\tau}_{x} &= \big(r \sin \theta \cos \phi, 1, 0 ,0 \big) \, ,  \\
	\hat{\tau}_{y} &= \big(r \sin \theta \sin \phi, 0, 1, 0 \big) \, , \\
	\hat{\tau}_{z} &= \big(r \cos \theta, 0, 0, 1\big) \, .
\end{align}
\end{subequations}
Rewriting these in the coincident gauge coordinates $\hat{x}^{\mu} = Y^{\mu}_{(2)} = (\tilde{t},x,y,z)$ leads to
\begin{subequations}
\begin{align}
	\tau_{x}(Y) &= ( x, 1,0,0 ) \, ,\\
	\tau_{y}(Y) &= (y, 0,1 ,0 ) \,,  \\
	\tau_{z}(Y) &= (z , 0, 0,  1) \, ,
\end{align}
\end{subequations}
which shows they are indeed affine in $Y^{\mu}_{(2)}$. Again, this coordinate transformation mixes space and time components in the translation Killing vectors. In the usual Cartesian form, the timelike components would instead be zero.
 Note that one can readily find the corresponding coordinate transformation between these two branches by simply composing the inverse of one transformation with the other $(Y_{(1)})^{-1} \circ Y_{(2)}$.
This new transformation is \textit{not} an affine transformation, so it is non-trivial that these separate branches exist. This is analogous to the so-called remnant group in metric teleparallel gravity~\cite{Ferraro:2014owa} (which is not a group), though there one is instead dealing with local Lorentz transformations and frame fields as opposed to coordinate transformations and the metric.

\item
Finally, we consider the case where both $T_{1}(t) =0$ and $\dot{R}(t)=0$, which automatically satisfies both differential equations~(\ref{Tdiffeq1Main}) and~(\ref{Tdiffeq2Main}).
The coordinate transformation is given by
\begin{equation}  \label{coordset3}
	\hat{x}^{0} = T_{2}(t) \, , \quad \hat{x}^{1} = r \sin \theta \cos \phi \, , \quad \hat{x}^{2} = r \sin \theta \sin \phi \, , \quad \hat{x}^{3} = r \cos \theta \, .
\end{equation}
This is simply the transformation to the standard Cartesian form of the flat FLRW metric along with a time reparametrisation $T_2(t)$. 
The transformed translation Killing vectors components are
\begin{subequations}
\begin{align}
	\hat{\tau}_{x} &= (0,1,0,0) \, ,  \\
	\hat{\tau}_{y} &= (0,0,1,0) \, , \\
	\hat{\tau}_{z} &= (0, 0, 0, 1) \, .
\end{align}
\end{subequations}
Note that they also take the same form in the coincident gauge coordinates because the components are constants and so immediately satisfy the coincident gauge constraints.

This branch can be easily obtained as a limit of the second branch~(\ref{coordset2}) by setting the constant coefficient of the $r^2$ term to zero instead of unity, $c_4 =0$. From the first branch, we must instead make the redefinition $R(t) = 1 + b_1 T_2 (t)$ and set the constants $c_2 = - c_1 = 1/b_1$ in~(\ref{T2sol}), such that the coordinate transformation is
\begin{equation}
	R(t,r) = r (1+b_1 T_2(t)) \, , \qquad T(t,r) = - \frac{1}{b_1}  + \frac{1+b_1 T_2(t)}{b_1} = T_2(t) \, ,
\end{equation}
after which we can take the limit $b_1 \rightarrow 0$. The same result is obtained in~\cite{DAmbrosio:2021pnd} in the covariant formalism, working instead with the affine connection coefficients.

\end{enumerate}

In summary, we have derived the three solutions for the flat FLRW coincident gauge coordinate transformations. These results agree with the previous results in the literature~\cite{Hohmann:2021ast,DAmbrosio:2021pnd}. However, the starting point and specific approach is different, as we do not explicitly work with the affine connection nor its symmetries. 

 It is also worth emphasising that these different coordinate systems represent unique solutions for the $f(\mathring{Q})$ theories, and hence the covariant $f(Q)$ theories as well. This is quite unusual compared to other gravitational theories, but makes complete sense when working with coordinate non-invariant models. The uniqueness of the solutions carries through to the covariant $f(Q)$ theories, where it is instead represented by different connection solutions as opposed to coordinate choices. One key difference between this work and previous studies in $f(Q)$ gravity is that now, \textit{all} geometric information will be encoded in the metric tensor, as opposed to split between the metric and the affine connection. This is a fundamental property of the symmetric teleparallel theories, not present in other metric-affine geometries.

\subsubsection{Spatially curved $k\neq0$ solutions}
For the spatially curved branch with $k \neq 0$ the situation is far simpler than in the previous case. In Appendix \ref{Append02} we find our partial solutions given in terms of just two free functions
\begin{equation} \label{partialSols2}
R(t,r) = r R(t) \, ,  \qquad T(t,r) = \sqrt{1-k r^2} T_1(t)  \, ,
\end{equation}
and only one remaining equation to be solved
	\begin{align}   \label{remainingkMain}
		R(t) \dot{T}_1(t) - T_1(t) \dot{R}(t) = 0 \, .
	\end{align} 
This equation simply gives $T_1 \propto R$, as we have $T_1 \neq 0$ and $R\neq0$ for the regularity of the coordinate transformation. Our final and only solution is then given by 
\begin{equation}
T(t,r) = \sqrt{1-k r^2} R(t)  \, , \qquad R(t,r) = r R(t)\, ,
\end{equation}
which satisfies $\hat{\partial}_{\mu} \hat{\partial}_{\nu} \hat{\tau}^{\lambda}_{i}=0$ for each of the translation generators $\hat{\tau}_{x}$, $\hat{\tau}_{y}$ and $\hat{\tau}_{z}$. For the spatially curved $k \neq 0$ branch the unique coordinate transformation is thus
\begin{align}  \label{coordsetk}
	\hat{x}^{0} = \sqrt{1-k r^2} R(t)\, , \quad \hat{x}^{1} = R(t) r \sin \theta \cos \phi \, , \quad \hat{x}^{2} = R(t) r \sin \theta \sin \phi \,  , \quad   \hat{x}^{3} = R(t) r \cos \theta \, .
\end{align}
The flat $k \rightarrow 0$ limit immediately reduces to first branch of transformation~(\ref{coordset1}), while to obtain the second branch one needs to make further redefinitions, see for instance~\cite{DAmbrosio:2021pnd}.

The translation Killing vectors have the following components
\begin{subequations}
\begin{align} \label{tauk}
	\hat{\tau}_{x} &= \big(-k R(t)  r \sin \theta \cos \phi, \sqrt{1-kr^2} R(t) , 0 ,0 \big) \, , \\
	\hat{\tau}_{y} &= \big(-k R(t) r \sin \theta \sin \phi, 0, \sqrt{1-kr^2} R(t) , 0 \big) \, , \\
	\hat{\tau}_{z} &= \big(-k R(t) r \cos \theta , 0, 0, \sqrt{1-kr^2} R(t) \big) \, ,
\end{align}
\end{subequations}
and in the coincident gauge coordinates $Y^{\mu}_{(k)} = (\tilde{t},x,y,z)$ this can be written in an affine form
\begin{subequations}
\begin{align}  \label{taucoinck}
	\tau_{x}(Y) &= \big( -k x,  \tilde{t} ,0,0 \big) \, ,\\
	\tau_{y}(Y) &= \big(-k y, 0,  \tilde{t} ,0 \big) \,,  \\
	\tau_{z}(Y) &= \big(-k z, 0, 0, \tilde{t}  \big) \, .
\end{align}
\end{subequations}
This result again agrees with the previous studies in symmetric teleparallel gravity~\cite{Hohmann:2021ast,Heisenberg:2022mbo}. But as we previously stated, this approach differs conceptually by not using the affine connection, highlighting the equivalence between the fully covariant treatment and the gauge-fixed approach. This also provides concrete evidence that claims that the coincident gauge is incompatible with spatial curvature $k\neq0$ are incorrect.

All four of the new sets of coordinates are adapted to the FLRW isometries in the sense that they transform the Killing vectors into affine functions (linear in the new coordinates). This is visualised in Appendix~\ref{Append1}, which gives some intuition on why they take the forms that they do. However, this intuition will always be limited by the fact that coordinates are not intrinsically physical. The choice of coordinates is usually based on convenience and expressing the metric in a simple form. On the other hand, here we have a well-defined and unambiguous method of selecting out particular sets of coordinates which satisfy the criteria of a vanishing affine connection. Again, this is reminiscent of the unitary gauge in massive gravity~\cite{deRham:2014zqa}.

As a last point, we note that within the general teleparallel gravity models, where both torsion and non-metricity are present, there are more branches of flat and curved solutions~\cite{Gomes:2023hyk}. There, it is also shown that the coincident gauge (or \textit{coincident frame}) is non-canonical. The additional branches of solutions in the general teleparallel geometries exist only in the canonical frames and therefore do not exist in the symmetric teleparallel theories. This is perhaps not too surprising, because allowing for torsion as well as non-metricity gives more freedom in the allowed space of solutions for the affine connection. Lastly, we emphasise that the calculations performed here apply to all symmetric teleparallel geometries and not just those in the $f(Q)$ theories.

\subsection{FLRW metric in coincident gauge coordinates}
With our four solutions for the coincident gauge coordinate transformations, we can now go ahead and state the form of the FLRW metric in these new coordinates. For clarity, we will label the four distinct coordinate sets as $Y^{\mu}_{(1)}$, $Y^{\mu}_{(2)}$, $Y^{\mu}_{(3)}$ and $Y^{\mu}_{(k)}$, referring to equations~(\ref{coordset1}), (\ref{coordset2}), (\ref{coordset3}), and~(\ref{coordsetk}) respectively. 

\subsubsection*{Flat branch 1}
 Beginning with the first solutions in the flat branch with coordinates $Y^{\mu}_{(1)}$,
\begin{equation} \label{coords1}
	\tilde{t} = R(t)  \, , \quad x = r R (t)   \sin \theta \cos \phi \, , \quad y = r R(t)  \sin \theta \sin \phi \, \qquad z = r R (t)  \cos \theta \, ,
\end{equation}
the metric tensor is given by 
\begin{equation} \label{met1}
g_{\mu \nu} (Y^{\mu}_{(1)}) = \begin{pmatrix}
\frac{ r^2  a^2}{\tilde{t}^2} - \frac{N^2}{\dot{R}^2}& -\frac{x a^2}{\tilde{t}^3} & -\frac{y a^2}{\tilde{t}^3} & -\frac{z a^2}{\tilde{t}^3}\\
 -\frac{x a^2}{\tilde{t}^3} &  \frac{a^2}{\tilde{t}^2} & 0 & 0  \\
  -\frac{y a^2}{\tilde{t}^3} & 0 &   \frac{a^2}{\tilde{t}^2} & 0  \\
    -\frac{z a^2}{\tilde{t}^3} & 0 &  0 &  \frac{a^2}{\tilde{t}^2}   \\ 
\end{pmatrix} \, ,
\end{equation}
where $r^2 = ( x^2 + y^2 +z^2 )/ \tilde{t}^2$  from Eq.~(\ref{coords1}), and the scale factor, lapse and transformation function $R$ are all defined in terms of the original (cosmological) time coordinate $t$. Assuming that the transformation $\tilde{t} = R(t)$ can be inverted, the functions can be written in terms of the coincident gauge coordinates $a(t) = a( R^{-1}(\tilde{t}))$, $N(t) = N(R^{-1}(\tilde{t}))$ and $\dot{R}(t) = R'(R^{-1}(\tilde{t}))$. This will be assumed going forward, but we will often use $t$ instead of $R^{-1}(\tilde{t})$ for notational ease.

The metric can be rewritten in a more compact form that is reminiscent of the 3+1 splitting by making use of the fluid 4-velocity in these coordinates. It is convenient to instead work first with the inverse metric
\begin{equation} \label{met1inv}
	g^{\mu \nu} = \begin{pmatrix}
		-\frac{\dot{R}^2}{N^2} & -\frac{x \dot{R}^2}{\tilde{t} N^2} & -\frac{y \dot{R}^2}{\tilde{t} N^2}  & -\frac{z
\dot{R}^2}{\tilde{t} N^2} \\
		-\frac{x \dot{R}^2}{\tilde{t} N^2}  & \frac{\tilde{t}^2}{a^2} -\frac{x^2 \dot{R}^2}{\tilde{t}^2 N^2} & -\frac{xy \dot{R}^2}{\tilde{t}^2 N^2}  & -\frac{xz \dot{R}^2}{\tilde{t}^2 N^2}   \\
		-\frac{y \dot{R}^2}{\tilde{t} N^2}  & -\frac{x y\dot{R}^2}{\tilde{t}^2 N^2}  & \frac{\tilde{t}^2}{a^2} -\frac{y^2 \dot{R}^2}{\tilde{t}^2 N^2}  & -\frac{yz \dot{R}^2}{\tilde{t}^2 N^2}   \\
		-\frac{z \dot{R}^2}{\tilde{t} N^2}  &-\frac{x z\dot{R}^2}{\tilde{t}^2 N^2}  &  -\frac{yz \dot{R}^2}{\tilde{t}^2 N^2}  &   \frac{\tilde{t}^2}{a^2} -\frac{z^2 \dot{R}^2}{\tilde{t}^2 N^2}    \\ 
	\end{pmatrix} \, ,
\end{equation}
which can then be expressed neatly as
\begin{equation} \label{met1comp}
	g^{\mu \nu} = -u^{\mu} u^{\nu} + \gamma^{\mu \nu} \, ,
\end{equation}
where the fluid 4-velocity $u^{\mu}$ and the spatial metric $\gamma^{\mu \nu}=\gamma^{ij}$ are given by
\begin{equation} \label{ugamma}
	u^{\mu} =- \frac{\dot{R}}{N} \left(1,\frac{x}{\tilde{t} },\frac{y}{\tilde{t}} , \frac{z }{\tilde{t} }  \right) \, , \quad \gamma^{i j} = \delta^{ij} \frac{\tilde{t}^2}{a^2} \, .
\end{equation}
The fluid 4-velocity is time-like normalised $u^{\mu} u_{\mu} = -1$ and can be easily found by making a coordinate transformation from the standard cosmological 4-velocity with components $U^{\mu} = (N,0,0,0)$.
In fact, $\gamma^{\mu \nu}$ is exactly the projection tensor with $\gamma^{\mu \nu} u_{\mu} = 0$ and $\gamma^2= \gamma$, which follow directly from~(\ref{met1comp}).

Under the transformation~(\ref{coordset1}), or working directly with the metric~(\ref{met1}), the 
 non-metricity scalar in the coincident gauge takes the form
\begin{equation}
\ourG \big(Y^{\mu}_{(1)} \big) = -\mathring{Q}_{(1)} =  - \frac{6H^2}{N^2} +  \frac{9H \dot{R} }{R N^2} + \frac{3}{N} \frac{d}{dt} \left( \frac{\dot{R}}{R N} \right) \,   \, .
\end{equation}
This can be simplified by making the field redefinition $K_1 = \dot{R}/(R N)$, where the fluid velocity becomes  $u^{\mu} = -K_1 \left( \tilde{t},x,y,z \right)$ and the non-metricity scalar can be written as
\begin{equation} \label{Q1}
 \mathring{Q}_{(1)} =  \frac{6H^2}{N^2} - \frac{9 H K_1}{N} - \frac{3 \dot{K_1}}{N} \, .
\end{equation}
We reserve the use of $\mathring{Q}$ for the non-metricity scalar in the coincident gauge coordinates, such that $\mathring{Q} \equiv \mathring{Q}(Y^{\mu})$, and so $\mathring{Q}_{(1)}$ is shorthand for $\mathring{Q}(Y^{\mu}_{(1)})$. As a result of Eq.~(\ref{G_decomp}), and the vanishing of the affine connection in this gauge, we have $Q=\mathring{Q}$. Hence, (\ref{Q1}) is equivalent to the full non-metricity scalar for one branch of solution in the spatially flat FLRW spacetime. Furthermore, because $Q$ is a true gauge-invariant scalar quantity, when one works in the covariant formalism of the symmetric teleparallel theories (with a non-vanishing connection and non-coincident coordinates), equation~(\ref{Q1}) remains the same. In that case, the (potential) degrees of freedom in $K_1(t)$ come from the non-vanishing components of the affine connection, instead of coming from additional free functions in the metric tensor. This is verified by comparison with ~\cite{Hohmann:2021ast,DAmbrosio:2021pnd}, which do not work in the coincident gauge but $Q$ takes the same form as~(\ref{Q1}), and similarly for the other branches below. In this section, we will continue to use the notation $\mathring{Q}$ to reinforce that we have derived this object from the coincident gauge metrics, but will drop the over-ring in the following section.

It is also then easy to verify if the coincident gauge has been correctly applied: the non-metricity scalar should be identical in the coincident gauge to in the full covariant formalism (working with a non-vanishing affine connection) for each given branch of solution. The same is true in reverse, working first with the coincident gauge objects and confirming the covariant formalism leads to the same results. For tensorial objects, it is less straightforward because they transform covariantly as opposed to invariantly. In that case, the explicit form of the coordinate transformation is needed to make a direct comparison.

\subsubsection*{Flat branch 2}
For the second flat branch with coordinates $Y^{\mu}_{(2)}$ and
\begin{equation}
	\tilde{t} = T (t) + \frac{r^2}{2}  \, , \quad x = r \sin \theta \cos \phi \, , \quad y = r \sin \theta \sin \phi \, , \quad z= r \cos \theta \, ,
\end{equation}
the metric tensor takes the form
\begin{equation} \label{met2}
g_{\mu \nu}(Y^{\mu}_{(2)}) = 
 \begin{pmatrix}
-\frac{N^2}{\dot{T}^2} & \frac{x N^2}{\dot{T}^2} & \frac{y N^2}{\dot{T}^2} &\frac{z N^2}{\dot{T}^2} \\
\frac{x N^2}{\dot{T}^2} & a^2 -\frac{x^2 N^2}{\dot{T}^2}& -\frac{xy N^2}{\dot{T}^2} & - \frac{xz N^2}{\dot{T}^2}  \\
\frac{y N^2}{\dot{T}^2} & -\frac{xy N^2}{\dot{T}^2} & a^2 -\frac{y^2 N^2}{\dot{T}^2} & -\frac{yz N^2}{\dot{T}^2} \\
\frac{z N^2}{\dot{T}^2} & -\frac{xz N^2}{\dot{T}^2} & -\frac{yz N^2}{\dot{T}^2} & a^2 -\frac{z^2 N^2}{\dot{T}^2} \\ 
\end{pmatrix} \, .
\end{equation}
We again assume the invertibility of the coordinate transformation such that $N$ and $\dot{T}$ are implicitly functions of the coincident gauge coordinates via $t = T^{-1}\big[\tilde{t} - (x^2+y^2+z^2)/2 \big]$. The matrix form of the metric components bears a resemblance to the inverse metric in Eq.~(\ref{met1inv}), and so we can similarly write this in a more compact form using a 3+1 style decomposition. Again, making use of the fluid 4-velocity $u^{\mu}$ in these coordinates and a spatial metric $\gamma_{\mu \nu} = \gamma_{ij}$, the metric can be written as 
\begin{equation} \label{met2a}
g_{\mu \nu}= - u_{\mu} u_{\nu} + \gamma_{\mu \nu} \, ,  \qquad u_{\mu} = \frac{N}{\dot{T}}\left(1,-x,-y,-z \right) \, ,  \qquad \gamma_{ij} =  \delta_{ij} a^2  \, ,
\end{equation}
where $u^{\mu}u_{\mu} =-1$. Keep in mind that the projection tensor $\gamma_{\mu \nu}$ for this metric is not equivalent to the projection tensor of the first spatially flat metric~(\ref{ugamma}), though they are closely related. Specifically, the covariant components of the former is equal to the contravariant components of the latter.

The gauge-fixed non-metricity scalar takes the form
\begin{equation}
\ourG \big(Y^{\mu}_{(2)} \big) =- \mathring{Q}_{(2)} = -\frac{6H^2}{N^2} +  \frac{9H}{\dot{T}} + \frac{3}{N} \frac{d}{dt} \left( \frac{N}{\dot{T} a^2}\right) \, .
\end{equation}
Again, we can make a field redefinition $K_{2} = N/(\dot{T} a^2)$, bringing this into the same form as the previous branch
\begin{equation} \label{Q2}
\mathring{Q}_{(2)}=  \frac{6H^2}{N^2} - \frac{9 H K_2}{N} - \frac{3 \dot{K_2}}{N}   \, .
\end{equation}
Note that $K_1 \neq K_2$ and hence the two non-metricity scalars are distinct from one another. The similarities will be further explored for the equations of motion.

\subsubsection*{Flat branch 3}
The third spatially flat branch $Y^{\mu}_{(3)}$ represents the standard spherical-to-Cartesian coordinate transformation, along with a reparametrisation of the time coordinate
\begin{equation}
	\tilde{t} = T (t)  \, , \quad x = r \sin \theta \cos \phi \, , \quad y = r \sin \theta \sin \phi \, , \quad z= r \cos \theta \, .
\end{equation}
The metric is then the usual FLRW metric with an additional time function 
\begin{equation} \label{met3}
g_{\mu \nu} \big(Y^{\mu}_{(3)}\big) = \textrm{diag} \left( -\frac{N^2}{\dot{T}},a^2,a^2,a^2 \right) \, ,
\end{equation}
and the coincident gauge non-metricity scalar in these coordinates is
\begin{equation}
\ourG  \big(Y^{\mu}_{(3)}\big)  =  - \mathring{Q}_{(3)} = - 6 \frac{H^2}{N^2} \, .
\end{equation}
In this case, the coordinate function is non-dynamical and plays no physical role\footnote{This can be verified explicitly from the field equations, see Section~\ref{sec:fe}.}. Cosmologies with this metric have been well studied and are especially simple due to the connection dropping out entirely. Moreover, despite the lack of diffeomorphism invariance, a residual time-reparametrisation invariance remains, such that the lapse can be set to one~\cite{BeltranJimenez:2019tme}. The dynamics are also known to be equivalent to the metric teleparallel $f(T)$ theories when working with the diagonal tetrad~\cite{Boehmer:2021aji}. This equivalence will be revisited in the following section.

\subsubsection*{Curved branch}
Finally, for the spatially curved solution with coordinates $Y^{\mu}_{(k)}$,
\begin{align}  \label{ktrans}
\tilde{t} =\sqrt{1-k r^2} R(t) \, , \quad x = R(t) r \sin \theta \cos \phi \, , \quad  y = R(t) r \sin \theta \sin \phi \,  , \quad   z= R(t) r \cos \theta \, ,
\end{align}
the metric takes a more complicated form
\begin{equation}
	\begin{split}
	g_{\mu \nu} \big(Y^{\mu}_{(k)} \big) =  &	\left(  
	\begin{matrix}
		\frac{a^2 \left(x^2+y^2+z^2\right)-\frac{N^2 s^2 \tilde{t}^2}{\dot{R}^2}}{s^4} & -\frac{x \tilde{t} \left(a^2+\frac{k N^2 s^2}{\dot{R}^2}\right)}{s^4} \\
		-\frac{x \tilde{t} \left(a^2+\frac{k N^2 s^2}{\dot{R}^2}\right)}{s^4} & \frac{a^2 \left(\tilde{t}^2+k \left(y^2+z^2\right)\right)-\frac{k^2 N^2 s^2
				x^2}{\dot{R}^2}}{s^4} \\
		-\frac{y \tilde{t} \left(a^2+\frac{k N^2 s^2}{\dot{R}^2}\right)}{s^4} & -\frac{k x y \left(a^2+\frac{k N^2 s^2}{\dot{R}^2}\right)}{s^4} \\
		-\frac{z \tilde{t} \left(a^2+\frac{k N^2 s^2}{\dot{R}^2}\right)}{s^4} & -\frac{k x z \left(a^2+\frac{k N^2 s^2}{\dot{R}^2}\right)}{s^4} \\
		\end{matrix} \right. \\
	 & \quad  \left.
	\begin{matrix}
		-\frac{y \tilde{t} \left(a^2+\frac{k N^2 s^2}{\dot{R}^2}\right)}{s^4} & -\frac{z \tilde{t} \left(a^2+\frac{k N^2 s^2}{\dot{R}^2}\right)}{s^4} \\
     - \frac{k x y \left(a^2+\frac{k N^2 s^2}{\dot{R}^2}\right)}{s^4} & -\frac{k x z \left(a^2+\frac{k N^2
				s^2}{\dot{R}^2}\right)}{s^4} \\
		\frac{a^2	\left(\tilde{t}^2+k \left(x^2+z^2\right)\right)-\frac{k^2 N^2 s^2 y^2}{\dot{R}^2}}{s^4} & -\frac{k y z \left(a^2+\frac{k N^2
				s^2}{\dot{R}^2}\right)}{s^4} \\
			-\frac{k y z\left(a^2+\frac{k N^2 s^2}{\dot{R}^2}\right)}{s^4} & \frac{a^2 \left(\tilde{t}^2+k \left(x^2+y^2\right)\right)-\frac{k^2 N^2 s^2 z^2}{\dot{R}^2}}{s^4}
	\end{matrix}
	 \right)
	 \end{split}
\end{equation}
where we have defined $s^2 = k(x^2+y^2+z^2) + \tilde{t}^2 = R(t)^2$.
This can be simplified considerably by introducing the convenient function
\begin{equation} \label{Bterm}
	B(t) := -\frac{k a(t)^2}{s^4} -\frac{k^2 N(t)^2}{s^2 \dot{R}(t)^2}  \, ,
\end{equation}
which is related to the (potential) new degrees of freedom residing in $\dot{R}(t)$, and then the metric reduces to
\begin{equation} \label{kmet}
g_{\mu \nu} \big(Y^{\mu}_{(k)} \big) = \begin{pmatrix}
\frac{B \tilde{t}^2}{k^2} + \frac{a^2}{k s^2} & \frac{B x \tilde{t}}{k}   & \frac{B y \tilde{t}}{k}  & \frac{B  z \tilde{t}}{k}   \\
 \frac{ B x \tilde{t}}{k}&  B x^2 + \frac{a^2}{s^2 } & B  xy  & B xz   \\
  \frac{B y \tilde{t}}{k} & B xy & B  y^2 + \frac{a^2}{ s^2}   & B  yz  \\
 \frac{B z \tilde{t}}{k} &B xz &B yz &B z^2 + \frac{a^2}{ s^2 }  \\ 
\end{pmatrix} \, .
\end{equation}
 Note that the limit $k \rightarrow 0$ is not immediately obvious in this form, but it does indeed lead back to the first metric~(\ref{met1}). This can be verified by expanding out all terms explicitly in terms of $k$, or instead rewriting it in a form closer to the decomposition~(\ref{met1comp}), which we shall do now.

Beginning with the inverse metric 
\begin{equation}
 g^{\mu \nu} = 
	\begin{pmatrix}
		\frac{k^2 \left(x^2+y^2+z^2\right)}{a^2}-\frac{\dot{R}^2 \tilde{t}^2}{N^2 s^2} & -x \tilde{t} \left(\frac{k}{a^2}+\frac{\dot{R}^2}{N^2 s^2}\right) & -y
		\tilde{t} \left(\frac{k}{a^2}+\frac{\dot{R}^2}{N^2 s^2}\right) & -z \tilde{t} \left(\frac{k}{a^2}+\frac{\dot{R}^2}{N^2 s^2}\right) \\
		-x \tilde{t} \left(\frac{k}{a^2}+\frac{\dot{R}^2}{N^2 s^2}\right) & \frac{\tilde{t}^2+k \left(y^2+z^2\right)}{a^2}-\frac{\dot{R}^2 x^2}{N^2 s^2} & -x y
		\left(\frac{k}{a^2}+\frac{\dot{R}^2}{N^2 s^2}\right) & -x z \left(\frac{k}{a^2}+\frac{\dot{R}^2}{N^2 s^2}\right) \\
		-y \tilde{t} \left(\frac{k}{a^2}+\frac{\dot{R}^2}{N^2 s^2}\right) & -x y \left(\frac{k}{a^2}+\frac{\dot{R}^2}{N^2 s^2}\right) & \frac{\tilde{t}^2+k
			\left(x^2+z^2\right)}{a^2}-\frac{\dot{R}^2 y^2}{N^2 s^2} & -y z \left(\frac{k}{a^2}+\frac{\dot{R}^2}{N^2 s^2}\right) \\
		-z \tilde{t} \left(\frac{k}{a^2}+\frac{\dot{R}^2}{N^2 s^2}\right) & -x z \left(\frac{k}{a^2}+\frac{\dot{R}^2}{N^2 s^2}\right) & -y z
		\left(\frac{k}{a^2}+\frac{\dot{R}^2}{N^2 s^2}\right) & \frac{\tilde{t}^2+k \left(x^2+y^2\right)}{a^2}-\frac{\dot{R}^2 z^2}{N^2 s^2} \\
	\end{pmatrix} \, , 
\end{equation}
and again using the fluid 4-velocity
\begin{equation}
	u^{\mu} = -\left(\frac{\tilde{t} \dot{R}}{s N}, \frac{x \dot{R}}{s N} , \frac{y \dot{R}}{s N}, \frac{z \dot{R}}{s N} \right) \, ,
\end{equation}
leads to the following expression
\begin{equation} \label{kmetsimp}
	g^{\mu \nu} = -u^{\mu} u^{\nu} + \gamma^{\mu \nu}_{(k)}\, ,  \qquad  \gamma^{\mu \nu}_{(k)} = \frac{s^2}{a^2}\textrm{diag}\left(k,1,1,1 \right)  - \frac{k}{a^2} x^{\mu} x^{\nu} \, ,
\end{equation}
where we define the vector $x^{\mu} = (\tilde{t},x,y,z)$.  Again $ \gamma^{\mu \nu}_{(k)}$ is the projection tensor for this metric, but it no longer contains only spatial components.
 In the $k \rightarrow 0$ limit, however, we have $s = \tilde{t}$ and the fluid 4-velocity and tensor $ \gamma^{\mu \nu}_{(k)}$ reduce to the spatially flat form given in~(\ref{ugamma}). Hence, the similarities with the first branch metric $g^{\mu \nu}(Y^{\mu}_{(1)})$ in Eq.~(\ref{met1comp}) is now immediately clear. Making field redefinitions such that the second branch is obtained in the flat limit should also be possible, but we do not explore this here.

All functions of cosmological time $t$ can be trivially rewritten in terms of the coincident gauge time using the inverse transformation $t= R^{-1}(s)$. Note that $s^2> 0$ follows from the requirement that $1 - kr^2 > 0$ for all $k$ in the original coordinates. For positive $k$ (and the $k \rightarrow 0$ limit), this imposes no restrictions on the new coordinates. For negative $k$, we instead have $\tilde{t}^2 > {|k| (x^2+y^2+z^2)}$, which is an artefact of the mixing of space and time coordinates in the transformation~(\ref{ktrans}). For reasonable choices of $R(t)$, the metric in these new coordinates covers the same physical space as the original coordinates, but it would be interesting to further investigate the global and asymptotic structure of these coordinates. In Appendix~\ref{Append1} we give a brief overview of these coordinates, showing the 2D $(\tilde{t},x)$-plane in terms of the original coordinates. This also demonstrates how these coordinates are adapted to the FLRW isometries for each of the different spatial curvatures.

For the gauge-fixed non-metricity scalar, we now have 
\begin{equation}
\ourG \big(Y^{\mu}{}_{(k)}\big) = - \mathring{Q}_{(k)} = -\frac{6H^2}{N^2} + \frac{9 H\dot{R}}{N^2 R} + \frac{3}{N} \frac{d}{dt}  \left(\frac{\dot{R}}{RN} \right)   + \frac{3k}{a^2} \left(2- \frac{H R}{\dot{R}} + \frac{N R^2}{\dot{R}^2} \frac{d}{dt} \left(\frac{\dot{R}}{RN} \right)  \right)  \, .
\end{equation}
Making the field redefinition $K = \dot{R}/(RN)$ allows us to write this in a simpler form
\begin{equation} \label{Qk}
 \mathring{Q}_{(k)} = \frac{3}{N^2} \left( 2 H^2 - 3H K N - N \dot{K} \right)  + \frac{3 k }{N a^2} \left(\frac{H}{K} -2N  - \frac{\dot{K}}{K^2} \right)\, ,
\end{equation}
which also matches that found in the literature~\cite{Hohmann:2021ast}. The fluid 4-velocity can then also be written in terms of $K$, making the form given in~(\ref{kmetsimp}) especially easy to work with.

From the forms of the metric in all of these branches, and the coincident gauge Killing vector fields of the previous section, one can verify that the Killing equation is indeed satisfied in these coordinates. Moreover, the Lie derivative of the gauge-fixed non-metricity scalars $\mathring{Q}$ with respect to each of the Killing vectors in its associated branch also vanishes, as expected.

On the topic of degrees of freedom, let us restate some of the previous discussions in a slightly different way:
in a fully diffeomorphism invariant theory, the additional functions $K_1(t)$, $K_2(t)$ or $K(t)$ could usually be removed via a coordinate transformation, specifically, the inverse of the ones given above. In that sense, they could be considered purely gauge degrees of freedom and non-physical. This is verified explicitly in Section~\ref{sec:fe}, where in the diffeomorphism invariant limit of the $f(\mathring{Q})$ theories (i.e., GR with $f(\mathring{Q}) = \mathring{Q} + \Lambda$), these $K$ terms drop out completely. Therefore any diffeomorphism invariant theory is blind to the differences in form of the metrics given above, as one would expect. This is non-trivial to show in general, as the task of proving that two different metrics describe the same physical spacetime is not straightforward and must be done on a case-by-case basis.

However, in a coordinate-dependent theory, such as the gauge-fixed formalism considered here, these additional functions can no longer be removed via coordinate transformations and therefore may represent physical degrees of freedom. An example that makes this point more clear is taking the Minkowski limit $a \rightarrow 1$, $k \rightarrow 0$. Clearly the metric then describes flat spacetime with a vanishing Riemann tensor, yet there are still additional free functions which cause $\mathring{Q}$ to be non-zero for certain branches. For instance, $\mathring{Q}_{(1)}$ in (\ref{Q1}), $\mathring{Q}_{(2)}$ in (\ref{Q2}) and $\mathring{Q}_{(k)}$ in (\ref{Qk}) all reduce to $-3\dot{K}$ which is non-zero. Interestingly, this implies that non-metricity does not vanish in the naive Minkowski limit because $\mathring{Q}_{\mu \nu \lambda} \neq 0$. In fact, this is in agreement with the results found in~\cite{Bahamonde:2022zgj}, where the Minkowski limit of spherically symmetric spacetimes has non-vanishing non-metricity. This may appear to be problematic from a physical standpoint; however, it has been pointed out that if matter is minimally coupled, it is blind to these terms and follows Levi-Civita geodesics~\cite{Bahamonde:2022zgj}. A more satisfactory argument would be that this naive Minkowski limit is incompatible with the coincident gauge -- i.e., that the Minkowski Killing vectors do not all satisfy the coordinate constraints in~(\ref{KVcoinc}).
It would be interesting to use the approaches developed in this work to further investigate the Minkowski limits of the symmetric teleparallel theories in the future.

Returning to the comparison with the covariant $f(Q)$ theories, these can be viewed as fully diffeomorphism invariant but with additional degrees of freedom from the symmetric teleparallel connection $\bar{\Gamma}^{\lambda}{}_{\mu \nu}$. These enter the dynamical equations via the non-metricity tensor $Q_{\mu \nu \lambda} = - \bar{\nabla}_{\mu} g_{\nu \lambda}$ and play the exact same role as our coordinate functions here. Emphasis should again be made that the physical content of both descriptions are the same.
It would appear that this type of argument can always be made for symmetry-breaking theories, massive gravity~\cite{deRham:2014zqa} and Ho\u{r}ava-Lifshitz gravity~\cite{Kimpton:2010xi} being key examples. In these other scenarios, this procedure has often been implemented by introducing a reference metric or connection. Nonetheless, though the presentation of the theory may differ, the degrees of freedom and their dynamics remain the same. This is also true in our case. However, the geometric interpretation of the Stueckelberg fields coming from the symmetry-restored Einstein action~(\ref{E_action}) as representing the affine connection of a non-Riemannian geometry is particularly appealing. More complicated geometric extensions in non-flat geometries $\bar{R}_{\mu \nu \lambda}{}^{\gamma} \neq 0$ could also be constructed, as outlined in~\cite{Boehmer:2023fyl}. This would seem to naturally lead to hybrid-Palatini-like theories~\cite{Tamanini:2012mi,Tamanini:2013ltp} (with an auxiliary teleparallel connection in addition to an independent metric-affine connection). For further details of these more general constructions and extended geometric theories, see~\cite{Jensko:2023lmn}.

In summary, in this section we have explicitly derived three distinct, spatially flat FLRW branches and one spatially curved FLRW branch for the symmetric teleparallel geometries. The metric and non-metricity scalar has then been display in these new coordinates. Notably, an affine connection is not needed as all of the dynamics are encoded explicitly in the new metric functions. The lack of diffeomorphism invariance once fixing the coincident gauge means that these new functions cannot be eliminated by coordinate transformations. Hence, these four different branches are physically distinct from one another and will in general lead to different observational signatures.
In the next section, we will explicitly study the dynamics of these different branches for the $f(Q)$ class of theories.

\section{Field equations and solutions}  \label{sec:fe}
To derive the metric field equations in the gauge-fixed formalism, one can simply take the equations of motion~(\ref{metric_EoM}) for the coincident gauge metrics derived above. Alternatively, one can use the standard FLRW metric in spherical coordinates~(\ref{FLRW}) along with the transformation rules for each of the non-covariant terms in the field equations, which can be found in~\cite{Boehmer:2021aji}.
These are both equivalent to taking the covariant $f(Q)$ field equations and fixing the gauge $\bar{\Gamma}=0$. We will verify that these different routes do indeed lead to the same equations of motion, matching the covariant $f(Q)$ field equations found previously.
    
As shown in the preceding section, the coordinate transformation for the FLRW metric led to four distinct branches: three for the spatially flat case and one for the curved case. Because of this, it is not possible to state the general form of the equations of motion unless one uses the more general coordinate transformations~(\ref{Anzatz}) with $R(t,r)$ and $T(t,r)$ not yet defined. However, this will not be in the coincident gauge form. Moreover, the general form of the equations are extremely unwieldy and not practical to state explicitly. We will instead look at the individual branches of the spatially flat $k=0$ solution and the spatially curved $k \neq 0$ solution. 

\subsection{Constraint equations} \label{sec:constr}
Beforehand, we wish to comment on the the connection field equation in the coincident gauge~(\ref{coincident_cons}), which we restate here
\begin{equation} \label{connection2}
\partial_{\mu} \partial_{\nu} (\sqrt{-g} \mathring{\mathscr{P}}^{\mu \nu}{}_{\lambda} f'(\mathring{Q}) ) = 0 \, .
\end{equation}
Recall that this equation could be derived in two different ways: the first, in the covariant $f(Q)$ framework, is to take the connection variations of the action and then fix the gauge $\bar{\nabla}_{\mu} = - \partial_{\mu}$; the second is to work in the gauge-fixed formalism and consider infinitesimal diffeomorphisms of the $f(\mathring{Q})$ action. Though both are equivalent, in the second approach it feels natural to demand that this constraint~(\ref{connection2}) hold off-shell, i.e., independent from the metric solutions, as a consistency condition. Furthermore, one could demand this be satisfied for \textit{all} models $f(\mathring{Q})$, which leads to constraints on the coordinate functions. In other words, we can use equation~(\ref{connection2}) to constrain the new (physical) degrees of freedom while staying agnostic about the underlying choice of model $f$. In this case, we will be using the FLRW symmetries and the previously derived metric tensors to constrain the new cosmological degrees of freedom.

Expanding the diffeomorphism constraint equation gives
\begin{multline} 
f^{(3)}(\mathring{Q})  \partial_{\mu} \mathring{Q} \partial_{\nu} \mathring{Q}  \mathring{\mathscr{P}}^{\mu \nu}{}_{\lambda}+  
 f''(\mathring{Q}) \left( \frac{2}{\sqrt{-g}} \partial_{\mu} \mathring{Q} \partial_{\nu} \left( \sqrt{-g}  \mathring{\mathscr{P}}^{\mu \nu}{}_{\lambda} \right) +  \mathring{\mathscr{P}}^{\mu \nu}{}_{\lambda} \partial_{\mu} \partial_{\nu} \mathring{Q}   \right) + \\
  f'(\mathring{Q}) 
\partial_{\mu} \partial_{\nu} (\sqrt{-g}  \mathring{\mathscr{P}}^{\mu \nu}{}_{\lambda})  = 0 \, , \quad 
\end{multline} 
where the terms proportional to $f'(\mathring{Q})$ vanish by virtue of the contracted Bianchi identity $\nabla_{\mu} G^{\mu}{}_{\nu} \equiv 0$, which can be equally expressed as ${\partial_{\mu} \partial_{\nu} (\sqrt{-g} \mathring{\mathscr{P}}^{\mu \nu}{}_{\lambda}) \equiv 0}$~\cite{Boehmer:2021aji}. For the above equation to hold for all choices of $f$, we obtain the following two conditions
\begin{align} \label{constr1}
 \partial_{\mu} \mathring{Q} \partial_{\nu} \mathring{Q}  \mathring{\mathscr{P}}^{\mu \nu}{}_{\lambda} &= 0  \, ,\\ 
\frac{2}{\sqrt{-g}} \partial_{\mu} \mathring{Q} \partial_{\nu} \left( \sqrt{-g} \mathring{\mathscr{P}}^{\mu \nu}{}_{\lambda} \right) + \mathring{\mathscr{P}}^{\mu \nu}{}_{\lambda} \partial_{\mu} \partial_{\nu} \mathring{Q} &= 0 \, . \label{constr2} 
\end{align}
Note that these equations are only obtained by explicitly assuming that the constraint equation~(\ref{connection2}) should satisfy two properties: firstly, it should vanish independently from the metric field equations; and secondly, hold for all choices of model $f$, such that the coefficients of both $f''(\mathring{Q})$ and $f^{(3)}(\mathring{Q})$ must vanish. Clearly if one specifies a particular model $f$ then other possible solutions exist. However, the power of this approach allows us to constrain the beyond-GR metric degrees of freedom in a fully general manner. It also will allow us to make a direct comparison with the $f(T)$ theories, which we explain below.

Emphasis should first be made that these equations~(\ref{constr1})--(\ref{constr2}) are coordinate-dependent and must be computed in the coincident gauge coordinates. Alternatively, one can use the spherical cosmological coordinates along with the transformation rules for $ \mathring{\mathscr{P}}$, which naturally follow from the definition in Eq.~(\ref{E}). If one does not follow this procedure correctly, working instead in non-coincident coordinates (but incorrectly assuming the vanishing of the affine connection), one will obtain erroneous results.  

This is the most general yet the most restrictive route to constraining the new free functions of the metric, allowing one to analyse the $f(Q)$ theories for all choices of $f$. This is reminiscent of the anti-symmetric $f(T)$ field equations.
  There, it has been noted that solutions with $T = \textrm{const.}$ reduce to GR with an effective cosmological constant term~\cite{Krssak:2018ywd}. The same is true in the metric teleparallel framework~\cite{DAmbrosio:2021pnd}, and one could study solutions of this form with $\mathring{Q} = \textrm{const}$ by fixing the new coordinate functions, automatically satisfying~(\ref{constr1})--(\ref{constr2}). It is not surprising that these effective GR solutions make the conservation equation vanish, as this is the diffeomorphism invariant limit of the $f(\mathring{Q})$ theories. The same result is obtained by directly setting $f \propto \mathring{Q}$ with $f'' \rightarrow 0$ in Eq.~(\ref{connection2}).
   Solutions beyond GR + $\Lambda$ (i.e., $\mathring{Q} \neq \textrm{const}$) require alternative methods, which in general are more complicated. 
  However, in the cosmological setting, a simple solution exists for the $k <0$ branch. The physical significance of solutions existing in cosmology which solves~(\ref{constr1})--(\ref{constr2}) for all choices of model $f$ will be discussed below.

Note that even in STEGR (i.e., the GR equivalent limit), where~(\ref{connection2}) is automatically satisfied, the above constraints~(\ref{constr1})--(\ref{constr2}) are actually non-trivial. However, with diffeomorphism invariance restored, these conditions no longer have the same meaning, as all coordinate choices are valid. One situation where this may still be beneficial is in reference to pseudotensors~\cite{BeltranJimenez:2019bnx}, but they do not play any role in the gravitational dynamics.
Importantly, we also note that for the flat branch with standard Cartesian coordinate $Y^{\mu}_{(3)}$, both constraints~(\ref{constr1}) and~(\ref{constr2}) are identically satisfied. Hence, there are no additional degrees of freedom associated with the new metric function $T(t)$ for any of the $f(Q)$ models for this branch. This verifies that $T(t)$ is non-dynamical in~(\ref{coordset3}).

Of particular interest are the other branches of solutions, which have recently begun to see more study in non-coincident gauge coordinates~\cite{Guzman:2024cwa,Dimakis:2022rkd,Subramaniam:2023okn}. As these other flat branches, corresponding to the coincident gauge metrics~(\ref{met1}) and~(\ref{met2}), had yet to be written down in such a form, we will state their equations of motion and verify that their dynamics are equivalent to those found in these recent works. We then move on to the spatially curved cases, with a focus on the $f(T)$ equivalence for negatively curved solutions.

\subsection{Zero spatial curvature $k=0$}
We begin with spatially flat branches $k=0$.
For the first solution in coordinates $Y^{\mu}_{(1)}$, we remind the reader that the metric~(\ref{met1}) and non-metricity scalar~(\ref{Q1}) could be written as
\begin{equation*} 
	g^{\mu \nu} = -u^{\mu} u^{\nu} + \gamma^{\mu \nu} \, , \qquad Q=  \frac{6H^2}{N^2} - \frac{9 H K_1}{N} - \frac{3 \dot{K_1}}{N}  \, .
\end{equation*}
The 4-velocity and spatial metric are given by $u^{\mu} = - K_1 (\tilde{t}, x, y,z ) $ and $\gamma^{\mu \nu}= \delta^{ij} \tilde{t}^2/a^2$, where we have introduced the function $K_1 = \dot{R}/(RN)$ with $\tilde{t} = R(t)$ coming from the coordinate transformation.
We then find that the diffeomorphism constraint equations~(\ref{constr1})--(\ref{constr2}) have a single independent component, which is proportional to the derivative of the non-metricity scalar~(\ref{Q1}),
\begin{equation} \label{conssol1}
	\frac{27 \left(4 H^2 \dot{N}+N \left(3 \dot{H} K_1 N+N \ddot{K}_1-\dot{K}_1 \dot{N}\right)+H N \left(-4
		\dot{H}+3 \dot{K}_1 N-3 K_1 \dot{N}\right)\right)^2}{N^8 \tilde{t}} = \frac{3 \dot{Q}^2}{N^2 \tilde{t}} = 0 \, ,
\end{equation}
where an overdot refers to the cosmological time derivative $d/dt$. Note that the over-ring on the non-metricity scalar has been dropped because $Q=\mathring{Q}$.
The solution to Eq.~(\ref{conssol1}) with $Q = \textrm{const.}$ reduces back to GR with an effective cosmological constant\footnote{Remember that here we are solving the diffeomorphism constraint equation while keeping $f$ unfixed. Looking at specific models for $f$ allows Eq.~(\ref{connection2}) to be solved without the stricter constraints of~(\ref{constr1}) and~(\ref{constr2}) individually.}, as will be explicitly shown below.
 For completeness, we display the full connection field equation in this compact form
\begin{align} \label{sol1con}
 	f'' \frac{3  \left( \ddot{Q} N + \dot{Q} \left(3 H N - \dot{N} \right) \right) }{ N^3 \tilde{t}} + f^{(3)} \frac{3 \dot{Q}^2}{ N^2 \tilde{t}} = 0 \, ,
\end{align}
where it is understood that $f=f(Q)$. This again vanishes for $\dot{Q} = 0$, as previously claimed. To derive this equation we have used the gauge-fixed conservation equation~(\ref{coincident_cons}), which is equivalent to the connection variations in the covariant approach.

The metric field equations have just two independent components
\begin{align} \label{sol1field1}
 	  f'' \frac{3 K_1 \dot{Q} }{N} + f' \left(\frac{6 H^2}{N^2}+Q\right)  -f &=2 \kappa  \rho  \, , \\
f'' \frac{\dot{Q} \left(4 H N -3 K_1 N^2\right)}{N^3}+ f' \frac{
	\left(6 H^2 N+4 \dot{H} N-4 H \dot{N}+N^3 Q\right)}{N^3}-f
		&=	-2 \kappa  p \, , \label{sol1field2}
\end{align}
which correspond to a modified Friedmann equation and acceleration equation respectively. Note, however, that the Friedmann equation is \textit{not} a constraint because of the $\dot{Q}$ terms being proportional to $\dot{H}$. We have also assumed a perfect fluid energy-momentum tensor 
\begin{equation}
	T_{\mu \nu} = \left( \rho + p \right) u_{\mu} u_{\nu} + g_{\mu \nu} p \, ,
\end{equation}
with normalised fluid 4-velocity $u^{\mu} u_{\mu} = -1$, given explicitly in~(\ref{ugamma}).
This matches the results found in the literature~\cite{DAmbrosio:2021pnd,Hohmann:2021ast,Guzman:2024cwa}\footnote{Note that in some works, such as~\cite{Guzman:2024cwa}, their function is defined as $f = F + Q$. Also bear in mind that their non-metricity scalar and non-metricity tensor is defined with an overall minus sign compared to ours, so some simple redefinitions are needed to show the field equations are indeed the same.}. The continuity equation follows from the metric field equations and the connection field equation
\begin{equation} \label{cont}
	\dot{\rho} + 3H \left( \rho +p\right) =0 \, .
\end{equation}
In the gauge-fixed formalism, this is a simple consequence of demanding the gravitational action be invariant under infinitesimal diffeomorphisms.
 
 If we use the solution for the conservation equation~(\ref{conssol1}) with $Q = Q_0 = \textrm{const.}$, the equations reduce to
\begin{align} \label{GReff1}
   \frac{6 H^2 f'(Q_0)}{N^2}  + f'(Q_0) Q_0 - f(Q_0 )&= 2 \kappa \rho \, , \\
    f'(Q_0)  \frac{ \left(6 H^2 N+4 \dot{H} N-4 H \dot{N}+N^3 Q_0 \right)}{N^3} - f(Q_0) &= - 2 \kappa p \, . \label{GReff2}
\end{align}
This explains the naming of these equations as a modified Friedmann and acceleration equation, despite the fact that the former is not a constraint in the full $f(Q)$ theories.
As is evident from both equations, this is simply General Relativity with an effective cosmological constant. This is made explicit with the redefinition $\Lambda_{\textrm{eff}} = f/f' - Q_0$ leading to
\begin{align} \label{eff1}
	\frac{6H^2}{N^2} - \Lambda_{\textrm{eff}} &= 2 \tilde{\kappa} \rho \, , \\
	 \frac{ \left(6 H^2 N+4 \dot{H} N-4 H \dot{N}\right)}{N^3} - \Lambda_{\textrm{eff}} &= - 2 \tilde{\kappa} p \label{eff2} \, ,
\end{align}
where $\kappa$ has been rescaled by the constant $f'(Q_0)$. For all following branches, the same limit with $Q = \textrm{const.}$ can always be found in principle, due to the a priori free functions in the non-metricity scalar. It has been pointed out that this differs from $f(T)$ gravity~\cite{DAmbrosio:2021pnd}, where the torsion scalar restricted by the FLRW symmetries does not have free functions capable of making it vanish; this follows from the symmetry reduction on the affine connection for FLRW isometries along with imposing vanishing total curvature and non-metricity. However, this is somewhat of a pathological solution, which could instead be obtained by simply fixing $f(T) = T + \Lambda$. Moreover, the $f(T)$ spin connection field equation\footnote{This is equivalent to the antisymmetric $f(T)$ field equation, and can be also viewed as a consistency condition~\cite{Jensko:2023lmn}.} is identically satisfied for all connection choices in cosmology, as explained in~\cite{Hohmann:2019nat}. Whereas in the $f(Q)$ case, we have additional constraints coming from the connection field equations that do not vanish identically. Nonetheless, in Section~\ref{subsec:k} we show how an equivalence with the $f(T)$ theories can emerge.

For the second branch $Y^{\mu}_{(2)}$, the metric~(\ref{met2}) and non-metricity scalar~(\ref{Q2}) take the form

\begin{equation*} 
	g_{\mu \nu}= - u_{\mu} u_{\nu} + \gamma_{\mu \nu} \, , \qquad  Q =  \frac{6H^2}{N^2} - \frac{9 H K_2}{N} - \frac{3 \dot{K_2}}{N} \, ,
\end{equation*}
where again the four-velocity can be written in terms of the function $K_2 = N/(\dot{T} a^2)$ as $u_{\mu} = - K_1 a^2 (-1, x ,y ,z)$, and the spatial metric has components $\gamma_{ij} = \delta_{ij} a^2$.
The connection equation is again proportional to the non-metricity and its time derivatives
		\begin{equation}
		 f''	\frac{a^7 K_2^2 \left(\dot{Q} \left(12 H^2-3 H K_2 N-3
				K_2 \dot{N}-2 N^2 Q\right)+3 K_2 N \ddot{Q}\right)}{N^3} +  f^{(3)} \frac{3 a^7 K_2^3 \dot{Q}^2}{N^2} = 0 \, .
		\end{equation}
Demanding this vanish for arbitrary $f$ enforces the consistency conditions~(\ref{constr1}) and~(\ref{constr2}), which once again leads to a constant $Q$. The analysis is the same as in the previous case, resulting in GR with an effective cosmological constant, Eqs.~(\ref{eff1})--(\ref{eff2}). The metric field equations are
\begin{align}
	- f'' \frac{3 K_2 \dot{Q} }{N}+f' \left(\frac{6 H^2}{N^2}+Q\right)-f &= 2 \kappa \rho \, , \\
	\frac{f'' N \dot{Q} (4 H-K_2 N)+f' \left(6 H^2 N+4 \dot{H} N-4 H \dot{N}+N^3 Q\right)}{N^3}-f &= - 2 \kappa p \, ,
\end{align}
and the continuity equation~(\ref{cont}) follows.
Again, these equations match what has been found previously in the literature~\cite{DAmbrosio:2021pnd,Hohmann:2021ast,Guzman:2024cwa}. It is interesting that they are equivalent in form to the first branch~(\ref{sol1field1})--(\ref{sol1field2}). However, the fields $K_{1}$ and $K_{2}$ have slightly different definitions, and their corresponding connection equations of motion also differ. Consequently, physical predictions will be different in general for these two sets of coincident gauge coordinates. For further details on the relationship between these two branches of solution, and on the parametrisations of $K_1$ and $K_2$, we refer to~\cite{Hohmann:2021ast}.

In the final branch $Y^{\mu}_{(3)}$, with the usual diagonal Cartesian metric~(\ref{met3}) and non-metricity scalar $Q = 6 H^2/N^2$, the connection equation of motion vanishes identically. The metric field equations then take their familiar form
\begin{align} \label{flateom1}
	f' \frac{12 H^2 }{N^2}- f &= 2 \kappa \rho \,, \\
	f'' \frac{48 H^2 \left(\dot{H} N-H \dot{N}\right)}{N^5}+ f' \frac{4 \left(\dot{H} N+H \left(3 H
		N-\dot{N}\right)\right) }{N^3}-f &= - 2\kappa p \, . \label{flateom2}
\end{align}
where the term $\dot{T}$ in the metric drops out entirely. This is the only branch that has been previously studied in the coincident gauge~\cite{BeltranJimenez:2019tme}.

It should be pointed out that for the first two flat branches studied above, $Y^{\mu}_{(1)}$ and $Y^{\mu}_{(2)}$, to solve the connection field equations for all $f$, the only solution is $Q = \textrm{const.}$, i.e., the somewhat trivial case that reduces the dynamics to GR + $\Lambda$. On the other hand, for the third branch $Y^{\mu}_{(3)}$ (the standard Cartesian one), the connection field equation is satisfied identically for all $f$. It is also known that in $f(T)$ gravity, the only flat solution is equivalent to this final branch $Y^{\mu}_{(3)}$. It would seem that this is no coincidence, as this is the only one without additional constraints coming from consistency conditions. This will be further discussed at in the following calculations, where the equivalence is less trivial.

\subsection{Non-zero spatial curvature $k \neq 0$} \label{subsec:k}

Moving on to the case with non-vanishing spatial curvature  $k \neq 0$, for the coincident gauge coordinates $Y^{\mu}_{(k)}$ the metric~(\ref{kmetsimp}) and non-metricity scalar~(\ref{Qk}) take the form
\begin{equation*} 
	g^{\mu \nu} = -u^{\mu} u^{\nu} + \gamma^{\mu \nu}_{(k)}\, , \qquad  Q = \frac{3}{N^2} \left( 2 H^2 - 3H K N - N \dot{K} \right)  + \frac{3 k }{N a^2} \left(\frac{H}{K} -2N  - \frac{\dot{K}}{K^2} \right)\, ,
\end{equation*}
with the vector $x^{\mu} = (\tilde{t},x,y,z)$, projection tensor $\gamma^{\mu \nu}_{(k)} = R^2/a^2 \textrm{diag}\left(k,1,1,1 \right)  - k x^{\mu} x^{\nu} /a^2$, fluid 4-velocity $u^{\mu} = -K x^{\mu}$, and coordinate function $K = \dot{R}/(RN)$. Also recall that $R(t)$ can be expressed in terms of the coincident gauge coordinates through $R^2(t) = k (x^2 + y^2 +z^2) + \tilde{t}^2$. With this more complicated metric, it is somewhat surprising that the field equations can still be written in a fairly accessible form. Even more remarkably, we will show that they can reduce to the $f(T)$ equations for the negatively spatially curved cosmologies.

Beginning with the constraint equations~(\ref{constr1}) and~(\ref{constr2}), we find only one independent component which is again proportional to the time derivative of the non-metricity scalar
\begin{equation} \label{constraint}
\frac{3 \tilde{t} \dot{Q}^2  \left(a^2 K^2+ k\right)}{a^2 K^2 N^2 R^2} = 0 \, .
\end{equation}
The full connection field equation is 
\begin{multline} \label{kconnec}
f'' \Bigg[
\left( 
\frac{ \tilde{t} \dot{Q}}{a^2 K^2 N^3 R \left(a^2 K^2+k\right)}
\right)
 \Bigg( 3 a^4 K^4 \left(\dot{N}-3 H N\right)-2 a^2 k K
	\left(-6 H^2+15 H K N-3 K \dot{N}+N^2 Q\right)  \\
   	+3 k^2 \left(N (H-4 K
	N)+\dot{N}\right) \Bigg)
	- \left( \frac{3 \tilde{t} 
	\ddot{Q} \left(a^2 K^2+k\right)}{a^2 K^2 N^2 R}\right) \Bigg] - 	f^{(3)} \frac{3 \tilde{t} \dot{Q}^2 
 \left(a^2 K^2+k\right)}{a^2 K^2 N^2 R^2} = 0 \, .
\end{multline}
Note that the terms multiplying $f''(Q)$ do vanish for any solutions to~(\ref{constraint}), even when $Q \neq \textrm{const.}$, but this is not obvious and involves some calculations to show explicitly.
The metric field equations take the form
\begin{align} \label{k_eom1}
	f'' \left(\frac{3 k \dot{Q}}{a^2 K N}+\frac{3 K \dot{Q}}{N}\right)+f'
	\left(\frac{6 k}{a^2}+\frac{6 H^2}{N^2}+Q\right)-f &= 2 \kappa \rho \, , \\
f'' \left(\frac{k \dot{Q}}{a^2 K N}+\frac{4 H \dot{Q}}{N^2}-\frac{3 K
	\dot{Q}}{N}\right)+f' \left(\frac{2 k}{a^2}+\frac{6 H^2}{N^2}-\frac{4 H
	\dot{N}}{N^3}+\frac{4 \dot{H}}{N^2}+Q\right)-f &= - 2 \kappa p \, , \label{k_eom2}
\end{align}
which should again be compared with~\cite{DAmbrosio:2021pnd,Hohmann:2021ast,Guzman:2024cwa}.
Along with connection equation~(\ref{kconnec}), they satisfy the continuity equation~(\ref{cont}). As expected, taking the limit $k \rightarrow 0$ leads precisely to the $Y^{\mu}_{(1)}$ branch, with the conservation equation and field equations reducing to~(\ref{conssol1})--(\ref{sol1field2}). Similarly, for the GR choice $f(Q) = Q + \Lambda$, all factors of $K$ drop out and one arrives at the standard cosmological field equations.
In the following analysis, we continue to assume $k \neq0$.

For these models, one immediately notices that the conservation constraints~(\ref{constraint}) can be solved without fixing $Q$ to be a constant. This represents genuine new models beyond GR for the whole class of $f(Q)$ theories. The solution is simply
\begin{equation} \label{K_sol}
	K(t) = \frac{\pm \sqrt{-k}}{a(t)} \, ,
\end{equation}
or in terms of $R(t)$,
\begin{equation} \label{R_sol}
	R(t) = R_0 \exp \left( \int^t \frac{ \pm \sqrt{-k} N(v)}{a(v)} dv \right) \, .
\end{equation}
These solutions are only real for $k < 0$, which rules out the positively curved spatial cosmologies. It is also no longer possible to take a well-defined limit $k \rightarrow 0$, as the coordinate transformation becomes singular in that case.
We will now continue with the solution~(\ref{K_sol}) for the negative spatial curvature cosmologies.

With $K(t)$ determined by~(\ref{K_sol}) and the constraint equation satisfied, we can return to the metric field equations. Let us introduce the real parameter $\delta$ satisfying $\delta^2 = 1$, in order to account for both positive and negative solutions for $K$. Equation~(\ref{K_sol}) is then written as $K = -\delta \sqrt{-k}/a$, where we have chosen the minus sign for future convenience when relating this to the $f(T)$ theories.
For this solution, the metric becomes considerably simpler due to the fact that the term $B(t)$ in Eq.~(\ref{Bterm}) reduces to
\begin{equation}
B(t) = \frac{(1-\delta^2) a^2}{\delta^2 R^4} = 0 \, ,
\end{equation}
which vanishes from $\delta^2 =1$. This then diagonalises the metric
\begin{equation}
g_{\mu \nu} (Y^{\mu}_{(k)}) = \textrm{diag} \left( \frac{a^2}{ k R^2},\frac{a^2}{R^2},\frac{a^2}{R^2},\frac{a^2}{R^2}\right) \, ,
\end{equation}
where $R(t)$ can be written using the solution~(\ref{R_sol}).
 The solution is valid for $k < 0$, and so the metric indeed has the correct signature.

Using this solution, the non-metricity scalar now takes the form
\begin{equation} \label{Q_solk}
Q =  6 \left( \frac{H}{N} + \frac{\delta \sqrt{-k}}{a} \right)^2 \, .
\end{equation}
The metric field equations then reduce to a modified Friedmann constraint 
\begin{equation}\label{Fried1}
f' \frac{12 H \left(a H + \delta  \sqrt{-k} N\right)}{a N^2} - f= 2 \kappa \rho \, ,
\end{equation} 
and a modified acceleration equation 
\begin{equation} \label{Accel}
- 
f''
\frac{4  \dot{Q} \left(a
	H+\delta  \sqrt{-k} N\right)}{a N^2}  
+ f'\left(-\frac{4 k}{a^2}+\frac{12 \delta  H \sqrt{-k}}{a N}+\frac{4 \left(3
   H^2+\dot{H}\right) N-2 H \dot{N}}{N^3}\right)- f
    =-\kappa p \, .
\end{equation}
 In the GR limit $f(Q) \propto Q$ all factors of $\delta$ vanish and the expected field equations emerge. This form is considerably simpler than the general equations in~(\ref{k_eom1})--(\ref{k_eom2}) and begins to look more like the Cartesian FLRW solution~(\ref{flateom1})--(\ref{flateom2}). This can be seen as a result of fixing the additional functions that were introduced from the coincident gauge coordinate transformation and the metric becoming diagonal for this solution.

Inspecting the dynamical equations reveals an exact equivalence with the analogous studies in $f(T)$ gravity. The torsion scalar is identical to our non-metricity scalar~(\ref{Q_solk}), and the $f(T)$ field equations are identical to our field equations\footnote{For the Weitzenb\"{o}ck case, compare Eq.~(22) and~(23) in~\cite{Coley:2022qug} to ours~(\ref{Fried1})--(\ref{Accel}) with lapse set to one $N(t)=1$. The former is identical (hence our choice of sign with $\delta$), and the latter only takes some basic manipulations to see equivalence. Alternatively, in the covariant formulation, compare with equations (5.3) in~\cite{DAmbrosio:2021pnd}, noting that their lapse is the square-root of ours. (Also note that there appears to be a typo in their torsion scalar, which contains no $\dot{a}$ terms.)}. This has long been known for the flat Cartesian FLRW metric~(\ref{met3}), but to our knowledge this has not been studied for any general case of the spatially curved FLRW cosmologies. Upon deeper inspection, this seems to reveal a more fundamental equivalence between the metric and symmetric teleparallel models. In the $f(T)$ case, the Weitzenb\"{o}ck gauge must be compatible with the vanishing of the anti-symmetric field equations, which can be viewed as a consistency constraint for the choice of tetrad coming from local Lorentz invariance, see also~\cite{Coley:2022aty}. For the $f(T)$ solution, e.g.,~\cite{Coley:2022qug}, they make no a priori assumptions about the choice of model $f$. This is also because the cosmological solutions identically satisfy the anti-symmetric field equations~\cite{Hohmann:2019nat}, regardless of the choice of $f(T)$.
 Here, our solution for the new degrees of freedom in the metric was also constrained by a symmetry requirement: diffeomorphisms (or equivalently, the connection field equation in the covariant version). We can also choose to make no restrictions on the model $f$, which we've now shown results in the exact same dynamical equations.

It seems clear that once consistency conditions have been implemented in the most general way possible, a deeper equivalence can be found between both types of theory. Notably, this implies that observations depending on background parameters will be unable to discern between $f(Q)$ and $f(T)$ gravity. This may be a natural consequence of the fact that both the non-metricity scalar $Q$ and torsion scalar $T$ can be constructed from different first-order decompositions of the Levi-Civita Ricci scalar~\cite{Jensko:2023lmn}. In the future, we will revisit this topic of studying the relationship between the metric and symmetric teleparallel theories, where an equivalence seems to emerge for the flat and negatively curved $k < 0$ cosmologies. In particular, we plan to extend this analysis beyond the background dynamics to include linear cosmological perturbations too, which is crucial for connecting with observations.

For the positive $k$ case, in order to have real solutions of~(\ref{constraint}), one is forced to instead solve $\dot{Q} = 0$. Comparing this with the case of $f(T)$ gravity, there does appear to be a physical distinction between the two classes of theories. For $k >0$, the solution for the torsion scalar is~\cite{DAmbrosio:2021pnd,Coley:2022qug}
\begin{equation} \label{Tkplus}
	T = 6\left( \frac{k}{a^2} - \frac{H^2}{N^2} \right) \, .
\end{equation}
Clearly if one tries to force $Q$ into this form, the constraint equation will be non-zero due to $\dot{Q} \neq 0$ and $a^2 K^2 + k \neq 0$. To be concrete, the function $K$ would take the form 
\begin{equation}
	K(t) = \frac{ K_0 \pm \sqrt{4k a^2 + K_0^2}}{2a^2} 	\, ,
\end{equation}
where $K_0$ is an integration constant,
yielding $Q = T$.  One then instead finds that the conservation constraint equation~(\ref{constraint}) does not vanish
\begin{equation}
  \frac{6 \tilde{t} \sqrt{ 4k a^4 + K_0^2}}{ N^2 \left( \sqrt{ 4k a^4 +K_0^2} \pm K_0 \right)} \dot{Q}^2 \neq 0 \, ,
\end{equation}
where clearly $\dot{Q} \neq 0$ from~(\ref{Tkplus}).

For positive values of $k$, there does not seem to be an obvious way to find a direct equivalence between the two modified teleparallel theories. Lastly, note that once the free function $K$ has been fully determined, $\dot{Q}=0$ cannot be solved without restricting the space of solutions. In other words, this would lead to constraints on $a(t)$ and $N(t)$ given by the differential equation
\begin{equation}
	\frac{k}{a^2} + \frac{\dot{H}}{N^2} - \frac{H\dot{N}}{N^3} =0 \, .
\end{equation}
Following this route would fix $Q$ to a constant but leads to inconsistencies, which is evident by comparing with Eqs.~(\ref{eff1})--(\ref{eff2}). The field equations would imply a constant scale factor and a constant transformation function $R(t)$, which in turn makes the coordinate system singular and the metric undefined. Hence, we do not consider these solutions any further.

It is somewhat surprising that when forcing the new degrees of freedom in the $f(Q)$ theories to satisfy the conservation equations coming from diffeomorphism invariance, this does not allow for solutions beyond GR for $k > 0$, nor for the first two spatially flat branches. Yet in $f(T)$ gravity, where one expects the cosmological equations to be more constrained (as vanishing non-metricity $Q_{\mu \nu \lambda}=0$ implies more constraints than vanishing torsion $T^{\lambda}{}_{\mu \nu}=0$), there seems to be more freedom. On the other hand, clearly if one lifts the restrictive constraint of solving the connection equation for arbitrary $f$, other non-GR solutions can be found for specific models.

It seems especially important to investigate whether other spacetimes are also heavily restricted in this gauge-fixed approach (if considering the conservation equation an off-shell consistency constraint for the coordinate transformation degrees of freedom). For spherically symmetric spacetimes, the coincident gauge was explored in~\cite{Bahamonde:2022zgj}, showing a very interesting array of coordinate solutions with new metric functions. Whether or not these coordinate sets and solutions can satisfy the consistency equations~(\ref{constr1})--(\ref{constr2}) for a general $f$ seems to yet be unexplored, except for the well-known case of Schwarzschild in isotropic coordinates which is identically satisfied~\cite{Boehmer:2021aji}. Lastly, it would be interesting to derive analogous conditions to our Eq.~(\ref{constr1})--(\ref{constr2}) for symmetric teleparallel theories beyond $f(Q)$ gravity, where the structure of these equations could differ significantly.
 
\section{Conclusions}  \label{sec:conc}

In this work we derived the coincident gauge coordinates for the spatially curved FLRW symmetries, using the Killing vector condition $\partial_{\mu} \partial_{\nu} \xi^{\lambda}=0$. For the spatially flat branch, three distinct solutions could be found, while the spatially curved branch exhibited only a single solution. This is in complete agreement with previous studied using the covariant approach, working explicitly with the symmetric teleparallel connection. The metric tensor in these coordinates no longer takes a simple form, but this is necessary if one wishes to fix the vanishing of the affine connection in the coincident gauge. This assumption has been made widely in the literature for many cases that are in fact not coincident gauge coordinates, so extra care must be taken here. If implemented consistently, however, the physics remains the same regardless of what gauge choice is made. We showed this explicitly for the $f(Q)$ theories, which have seen the most attention recently.

The methodologies proposed here can now be applied in a wide range of different contexts. For instance, any studies in symmetric teleparallel gravity must ensure that either a non-vanishing teleparallel connection is used (and that it is consistent with spacetime symmetries), or that the Killing vectors satisfy the constraint~(\ref{condKV}). The procedures of Section~\ref{sec:kvf} can then be followed to find the correct coordinate transformations and investigate the new coincident gauge metrics. This is vital for a correct analysis and understanding of the theory in question.

Interestingly, if one forces the gauge-fixed action to be invariant under diffeomorphisms, one is led to the symmetric teleparallel connection field equation (fixed to the coincident gauge). We then explored the possibility of considering this as an \textit{off-shell} constraint; this severely restricts the new degrees of freedom but results in a direct equivalence with the $f(T)$ gravity theories for flat and negative spatially curved cosmologies. For the positive spatially curved FLRW metric, the constraints trivially reduce $f(Q)$ to GR with an effective cosmological constant, which is of course also included in the $f(T)$ models.
 These equivalences warrant further investigation, and may be relevant when assessing the theoretical viability of both $f(Q)$ and $f(T)$ gravity --- the former known to exhibit ghost pathologies and the latter facing issues of strong coupling. This does not seem to be a coincidence, as both theories are intimately related by their decompositions of the Ricci scalar, one in the coordinate basis and the other in the orthonormal basis.

We did not consider cosmological perturbations in this approach, which has been studied thoroughly in the covariant formalism for all branches of symmetric teleparallel gravity~\cite{Heisenberg:2023wgk}. The covariant approach seems to be the most practical way to study perturbations, as one is still able to exploit gauge invariance in the usual way. However, just as we have shown for the background dynamics, the end results should again be the same in either formalism. On the other hand, it would be very interesting to investigate the cosmological perturbations of the conservation equation derived in Section~\ref{sec:constr}. In this work, we showed that solving this constraint led to an equivalence with $f(T)$ gravity at the background level. The natural extension is to investigate whether linear perturbations are also equivalent. This is especially important, as it will determine whether observations can or cannot distinguish between either class of theory.  Such a project is outside of the scope of this work, but we plan to investigate this topic in the future.

Lastly, we point out that the coordinates derived in this work are applicable to all theories within the symmetric teleparallel geometry, not just the $f(Q)$ modifications. This is especially important in light of the aforementioned studies that indicate $f(Q)$ to be pathological~\cite{Gomes:2023tur}, with unstable ghost degrees of freedom present around cosmological backgrounds. However, it has also recently been shown in~\cite{Bello-Morales:2024vqk} that symmetric teleparallel modifications which are constructed to break diffeomorphisms but leave transverse diffeomorphisms unbroken are in fact ghost-free, see also~\cite{Alvarez:2006uu}. This appears to be a viable route of constructing healthy gravitational models beyond GR within these geometries. Again, our construction to find coincident gauge coordinates will be applicable to these theories too, as well as in more general modifications and extensions working within this geometry.

\subsection*{Acknowledgements}
The author thanks Christian Boehmer \& Tomi Koivisto for helpful comments and discussions.
This work was supported by the Engineering and Physical Sciences Research Council [EP/W524335/1].

\appendix
\addtocontents{toc}{\protect\setcounter{tocdepth}{0}}

\section{Coincident gauge translation Killing vectors} \label{Append0} 
Solving the translation Killing vector equations~(\ref{t_eqs}) can be done systematically using Mathematica, working through the components of $\hat{\tau}^{\lambda}_{i}$. The set of 64 equations coming from each $\hat{\partial}_{\mu} \hat{\partial}_{\nu} \hat{\tau}^{\lambda}_{i}=0$, after making various substitutions and manipulations, lead to differential equations that can be solved explicitly. We will study the spatially flat $k=0$ and spatially curved $k \neq 0$ cases separately, showing that the former has three distinct branches of solution.

\subsection{$k=0$}  \label{Append01}
We start by fixing $k=0$ and examining the equations $\hat{\partial}_{\mu} \hat{\partial}_{\nu} \hat{\tau}^{1}_{x}=0$. The general method is to eliminate higher derivatives of $T(t,r)$ and $R(t,r)$ until we are left with exactly solvable equations. 
Beginning with the set of four equations $\hat{\partial}_{t} \hat{\partial}_{\mu} \hat{\tau}^{1}_{x}$, solving $\hat{\partial}_{t} \hat{\partial}_{t} \hat{\tau}^{1}_{x}=0$ for $\partial_{r} \partial_{r} \partial_{r} T$ and $\hat{\partial}_{t} \hat{\partial}_{\phi} \hat{\tau}^{1}_{x}=0$ for $\partial_{t} \partial_{t} \partial_{r} T$ allows us to eliminate the third derivatives of the functions $T$ and $R$ in the following equation
\begin{equation} \label{diffT1}
	 \hat{\partial}_{t} \hat{\partial}_{r} \hat{\tau}^{1}_{x} = \partial_r \partial_r T \partial_t R - \partial_t \partial_r T \partial_r R = 0 \, .
\end{equation}
Rearranging this expression and taking various derivatives with respect to $r$ or $t$ then solves all four equations $\hat{\partial}_{t} \hat{\partial}_{\mu} \hat{\tau}^{1}_{x}=0$.

We now move on to a different set of equations $\hat{\partial}_{r} \hat{\partial}_{\mu} \hat{\tau}^{1}_{x} = 0$, making use of the previous solutions. The first expression $\hat{\partial}_{r} \hat{\partial}_{t} \hat{\tau}^{1}_{x}$ vanishes, and so we rearrange $\hat{\partial}_{r} \hat{\partial}_{r} \hat{\tau}^{1}_{x}=0$ for $\partial_{t} \partial_{r} \partial_{r} T$. Along with the previous substitutions for $\partial_{r} \partial_{r} \partial_{r} T$ and  $\partial_{t} \partial_{t} \partial_{r} T$, and obtaining $\partial_{t} \partial_{r} T$ from~(\ref{diffT1}), there is then only one remaining independent equation for the $ \hat{\tau}^{1}_{x}$ set of equations,
\begin{equation} \label{diffeq1}
\hat{\partial}_{r} \hat{\partial}_{\theta} \hat{\tau}^{1}_{x}= \partial_{r} R \partial_{r} T - R \partial_{r} \partial_{r} T = 0 \, .
\end{equation}
The two differential equations~(\ref{diffT1}) and~(\ref{diffeq1}) completely solve $\hat{\partial}_{\mu} \hat{\partial}_{\nu} \hat{\tau}^{1}_{x}=0$, and we can now move on to different components of the $x$-translation Killing vectors.
Recall from our original coordinate transformation~(\ref{Anzatz}) that these new functions cannot be zero, $R \neq 0$ and $T \neq 0$, or else the transformation is undefined. However, $T$ being a function of $t$ alone with $\partial_{r} T = 0$ is a perfectly valid solution. Hence, naively integrating~(\ref{diffeq1}) to obtain $R(t,r) = R_1(t) \partial_{r} T(t,r)$ would be incompatible with solutions with $\partial_{r} T = 0$. For this reason, we do not attempt to solve these expressions at this stage.

For the next set of equations, let us work with the fourth component of the $x$-translation Killing vectors $\hat{\partial}_{\mu} \hat{\partial}_{\nu} \hat{\tau}^{4}_{x}=0$. Repeating the process of eliminating higher derivatives, we solve the equation $\hat{\partial}_{t} \hat{\partial}_{t} \hat{\tau}^{4}_{x}=0$  for $\partial_{r} \partial_{r} \partial_{r} R$, the equation $\hat{\partial}_{t} \hat{\partial}_{\phi} \hat{\tau}^{4}_{x}=0$  for $\partial_{t} \partial_{t} \partial_{r} R$, the equation $\hat{\partial}_{\phi} \hat{\partial}_{\phi} \hat{\tau}^{4}_{x}=0$  for $\partial_{t} \partial_{r} \partial_{r} R$ and finally the equation $\hat{\partial}_{\theta} \hat{\partial}_{\theta} \hat{\tau}^{4}_{x}=0$ for $\partial_{r} \partial_{r} R$. We are then left with the following equation
\begin{equation}  \label{diff_solv}
 \hat{\partial}_{\phi} \hat{\partial}_{r} \hat{\tau}^{4}_{x} =  r \partial_{r} R - R = 0 \, ,
\end{equation}
which we solve to obtain
\begin{equation}
	R(t,r) = r R(t) \, ,
\end{equation}
where $R(t)$ now depends only on time $t$ and cannot vanish. This turns out to be the only independent equation remaining, and the solution satisfies all 16 equations in $\hat{\partial}_{\mu} \hat{\partial}_{\nu} \hat{\tau}^{4}_{x}=0$.
Plugging this solution back into the differential equation~(\ref{diffeq1}) yields
\begin{equation} \label{Teqtemp}
	\partial_{r}T - r \partial_{r} \partial_{r}T = 0\, ,
\end{equation}
which we easily solve for $T$
\begin{equation} \label{Tdiff2}
	T(t,r) = \frac{1}{2} r^2 T_{1}(t) + T_{2}(t) \, .
\end{equation}

Finally, returning to our initial equation~(\ref{diffT1}), this becomes
\begin{equation}  \label{Tdiffeq1}
 	T_1(t) \dot{R}(t) - R(t) \dot{T}_1(t) = 0 \, ,
\end{equation}
whose solutions solve all components of $\hat{\partial}_{\mu} \hat{\partial}_{\nu} \hat{\tau}^{\lambda}_{x}=0$ except for $\hat{\tau}^{2}_x$.
We therefore repeat the process for $\hat{\partial}_{\mu} \hat{\partial}_{\nu} \hat{\tau}^{2}_{x}$,
where we find 
\begin{equation} \label{Tdifftemp}
	\hat{\partial}_{t} \hat{\partial}_{t} \hat{\tau}^{2}_{x}= r^2 T_{1} \dot{R}^3 + R^2 \left(\dot{R} \ddot{T}_2 - \dot{T}_{2} \ddot{R} \right) =
	0 \, .
\end{equation}
Substitution back into $\hat{\partial}_{\mu} \hat{\partial}_{\nu} \hat{\tau}^{2}_{x}=0$ leaves the single the condition 
\begin{equation} \label{Rdiff}
  	T_1(t) \dot{R}(t) = 0 \, ,
\end{equation}
from which we also have $R(t) \dot{T}_1 = 0$ from~(\ref{Tdiffeq1}).
Clearly the first term in the previous equation~(\ref{Tdifftemp}) must vanish for any solutions of~(\ref{Rdiff}), so we are left with
\begin{equation} \label{Tdiffeq2}
	\dot{T}_2(t) \ddot{R}(t) - \dot{R}(t) \ddot{T}_2(t) = 0 \, .
\end{equation}
Solutions to these three equations~(\ref{Tdiffeq1}), (\ref{Rdiff}) and~(\ref{Tdiffeq2}) then solve all three sets of the 64 components of $\hat{\partial}_{\mu} \hat{\partial}_{\nu} \hat{\tau}^{\lambda}_{i}=0$ for each of the translation generators $\hat{\tau}_{x}$, $\hat{\tau}_{y}$ and $\hat{\tau}_{z}$. The partial solutions and remaining differential equations are given in the main body of text in equations~(\ref{partialSols1})-(\ref{Tdiffeq2Main}).

\subsection{$k \neq 0$}   \label{Append02}
For the spatially curved branch with $k \neq 0$, we repeat the exact same steps as before: after using equations $\hat{\partial}_{t} \hat{\partial}_{t} \hat{\tau}^{1}_{x}=0$ and  $\hat{\partial}_{t} \hat{\partial}_{\phi} \hat{\tau}^{1}_{x}=0$ to eliminate $\partial_{r} \partial_{r} \partial_{r} T$ and  $\partial_{t} \partial_{t} \partial_{r} T$ we arrive at an analogous differential equation to~(\ref{diffT1}) but with factors of $k$,
\begin{equation} \label{diffk1}
	\hat{\partial}_{t} \hat{\partial}_{r} \hat{\tau}^{1}_{x} = 
	(1-k r^2)
	\left( \partial_r \partial_r T \partial_t R - \partial_t \partial_r T \partial_r R  \right) - k r \partial_{r}T \partial_{t}R = 0 \, ,
\end{equation}
which reduces to Eq.~(\ref{diffT1}) when $k \rightarrow 0$. This then solves all $\hat{\partial}_{t} \hat{\partial}_{\mu} \hat{\tau}^{1}_{x}=0$. Moving on to $\hat{\partial}_{r} \hat{\partial}_{\mu} \hat{\tau}^{1}_{x}=0$,
and making the same substitutions as we did to obtain~(\ref{diffeq1}), i.e., using $\hat{\partial}_{r} \hat{\partial}_{r} \hat{\tau}^{1}_{x}=0$ to remove $\partial_{t} \partial_{r} \partial_{r} T$, along with the previous substitutions, 
we arrive at
\begin{equation} \label{diffk2}
	\hat{\partial}_{r} \hat{\partial}_{\theta} \hat{\tau}^{1}_{x}=
	(1-kr^2) \partial_{r} R \partial_{r} T + R \left(k r \partial_{r} T - (1-kr^2) \partial_{r} \partial_{r} T \right) 
	= 0 \, .
\end{equation}
Again, this should be compared with~(\ref{diffeq1}), which this equation reduces to for $k \rightarrow0$.

Looking at the fourth component of the generator $\hat{\tau}_{x}^4$, we now eliminate $R$ and its derivatives, making substitutions for $\partial_{r} \partial_{r} \partial_{r} R$, $\partial_{t} \partial_{t} \partial_{r} R$,  $\partial_{t} \partial_{r} \partial_{r} R$ and  $\partial_{r} \partial_{r} R$ from equations $\hat{\partial}_{t} \hat{\partial}_{t} \hat{\tau}^{4}_{x}=0$, $\hat{\partial}_{t} \hat{\partial}_{\phi} \hat{\tau}^{4}_{x}=0$, $\hat{\partial}_{\phi} \hat{\partial}_{\phi} \hat{\tau}^{4}_{x}=0$ and $\hat{\partial}_{\theta} \hat{\partial}_{\theta} \hat{\tau}^{4}_{x}=0$ respectively. With these substitutions we obtain exactly Eq.~(\ref{diff_solv}), so again have 
	\begin{equation}
		R(t,r) = r R(t) \,  ,
	\end{equation}
which satisfies all 16 equations in  $\hat{\partial}_{\mu} \hat{\partial}_{\nu} \hat{\tau}^{4}_{x}=0$.
For this solution, $R(t)$ drops out completely from~(\ref{diffk2}), and it reduces to 
\begin{equation}
	\partial_{r} T - r (1-k r^2) \partial_{r}\partial_{r} T = 0 \, .
\end{equation}
which is the analogue of~(\ref{Teqtemp}) with nonzero $k$. Solving this explicitly leads to 
\begin{equation}
	T(t,r) = -\frac{\sqrt{1-k r^2}}{k} T_1(t) + T_2(t) \, ,
\end{equation}
where we can no longer obtain the flat solution~(\ref{Tdiff2}) in the direct $k \rightarrow0$ limit. We will absorb the factor of $-1/k$ into the $T_1(t)$ term such that the solution can be written as $T(t,r) = \sqrt{1-k r^2} T_1(t) + T_2(t) $. Returning to our first equation~(\ref{diffk1}), this now takes the form
\begin{equation} \label{remainingk}
	R(t) \dot{T}_1(t) - T_1(t) \dot{R}(t) = 0 \, ,
\end{equation}
which again matches~(\ref{Tdiffeq1}). 

Looking now at the $\hat{\partial}_{\mu} \hat{\partial}_{\nu} \hat{\tau}^{2}_{x}=0$ equations, we first find
\begin{equation}
	\hat{\partial}_{t} \hat{\partial}_{t} \hat{\tau}^{2}_{x} = -k r^2 \dot{R}^2 \dot{T}_2 + (1-kr^2) R \dot{T}_2 \ddot{R} - (1-kr^2) R \dot{R} \ddot{T}_2 = 0 \, .
\end{equation}
Substitution of this back into the other non-vanishing components of $\hat{\partial}_{\mu} \hat{\partial}_{\nu} \hat{\tau}^{2}_{x}=0$  leads to another departure from the flat case, where instead of Eq.~(\ref{Rdiff}) we obtain two independent equations 
\begin{align} \label{appendfinal1}
	\hat{\partial}_{t} \hat{\partial}_{\theta} \hat{\tau}^{2}_{x} &= \frac{k \dot{T}_2 \dot{R} \left( \sqrt{1-kr^2} T_1 \dot{R} + (1-kr^2) R \dot{T}_2 \right)}{\left(T_1 \dot{R} + \sqrt{1-kr^2} R \dot{T}_2 \right)^3 } = 0 \, , \\
	\hat{\partial}_{r} \hat{\partial}_{\theta} \hat{\tau}^{2}_{x} &= \frac{k \dot{T}_2 \left(T_1^2 \dot{R}^2 - (1- kr^2) R^2 \dot{T}_2^2\right)}{\left(T_1 \dot{R} + \sqrt{1-kr^2} R \dot{T}_2 \right)^3 } = 0 \, .  \label{appendfinal2}
\end{align}
The term in the numerator bracket of~(\ref{appendfinal1}) cannot vanish or else the denominator of both equations also vanishes. Similarly, if the numerator bracket of~(\ref{appendfinal2}) vanishes then equation~(\ref{appendfinal1}) does not possess real solutions. Lastly, if $\dot{R}=0$ then equation (\ref{appendfinal1}) is satisfied but equation~(\ref{appendfinal2}) is non-zero. Therefore the only remaining possibility is that $\dot{T}_2=0$ with $T_2(t) = c$, which we set to zero without loss of generality. With this solution, we find out final transformation function to be 
\begin{equation}
T(t,r) = \sqrt{1-k r^2} T_1(t) \, ,
\end{equation}
along with the remaining equation~(\ref{remainingk}), as presented in Eqs.~(\ref{partialSols2})-(\ref{remainingkMain}) of the main text. These then satisfy all components of $\hat{\partial}_{\mu} \hat{\partial}_{\nu} \hat{\tau}^{\lambda}_{i}=0$ for each of the translation generators $\hat{\tau}_{x}$, $\hat{\tau}_{y}$ and $\hat{\tau}_{z}$.

\section{Geometric interpretation of coincident gauge coordinate systems}
\label{Append1}
The coincident gauge coordinates are adapted to the Killing vectors, such that they take an affine form $\xi^{\mu} = a^{\mu}{}_{\nu} \xi^{\nu} + b^{\mu}$ in the new coordinates, satisfying the condition $\partial_{\mu} \partial_{\nu} \xi^{\lambda} = 0$ in Eq.~(\ref{condKV}). 
 Note that this is slightly different to the usual adaptation of a coordinate system to a set of given Killing vectors, which uses coordinates where the Killing vectors form a basis and the coordinate lines $\hat{x}^{\mu}$ are Killing trajectories.

As a prime example, let us take the spatially curved FLRW coordinates given in~(\ref{ktrans}). For simplicity, let us just examine the $t$ and $r$ coordinates by fixing $\theta = \pi/2$ and $\phi=0$ such that $y=z=0$. We then have 
\begin{equation} \label{A_trans}
	\tilde{t} = \sqrt{1-k r^2} R(t) \, , \qquad x = r R(t) \, ,
\end{equation}
with the inverse transformations given by
\begin{equation} \label{tr}
	t = R^{-1}\left(\sqrt{k x^2 + \tilde{t}^2} \right)  \, , \qquad r = \frac{x}{\sqrt{k x^2 + \tilde{t}^2}} \, .
\end{equation}
From the initial FLRW metric we have $1-kr^2 > 0$, which translates to
\begin{equation*}
	\frac{\tilde{t}^2}{k x^2 + \tilde{t}^2} > 0 \ \implies \  k x^2 +  \tilde{t}^2 > 0 \, .
\end{equation*}
This is just the 2D version of our previous statement $s^2 >0$, discussed below Eq.~(\ref{kmetsimp}). This constrains the new coordinates when $k$ is negative.

\begin{figure}[bp!]
	\centering
	\begin{subfigure}[b]{0.73\textwidth}
		\includegraphics[width=\textwidth]{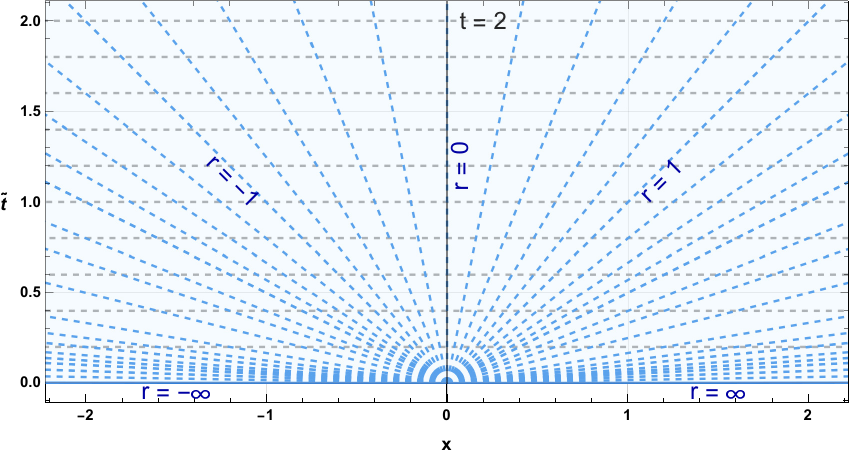}
		\caption{\centering{$k=0$}}
		\label{fig:A}
	\end{subfigure} 
	\begin{subfigure}[b]{0.73\textwidth}
		\includegraphics[width=\textwidth]{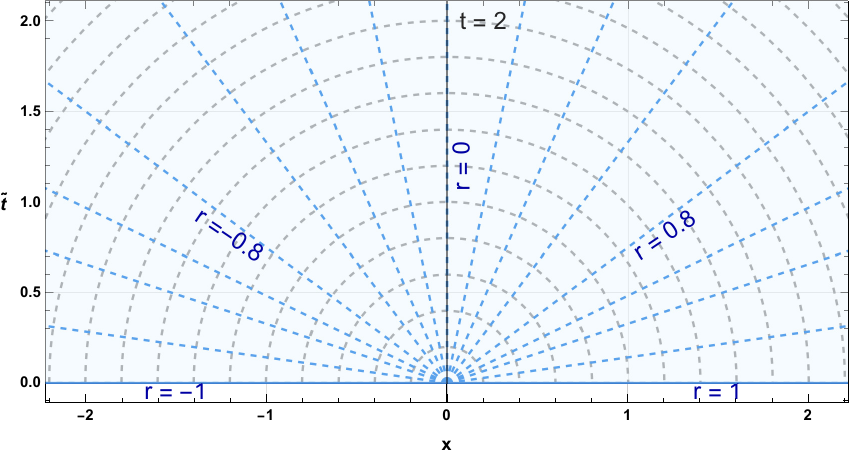}
		\caption{\centering{$k=1$}}
		\label{fig:B}
	\end{subfigure} 
	\begin{subfigure}[b]{0.73\textwidth}
		\includegraphics[width=\textwidth]{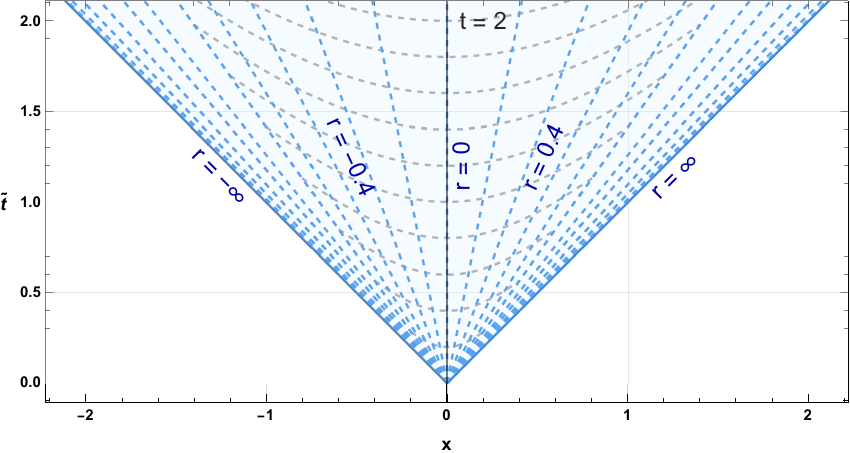}
		\caption{\centering{$k=-1$}}
		\label{fig:C}
	\end{subfigure} 
	\caption{Representation of coordinate transformation~(\ref{A_trans}) from cosmological coordinates $(t,r)$ to coincident gauge coordinates $(\tilde{t},x)$ for the choice $R(t)=t$. The dashed grey lines are lines of constant $t$, while the dashed blue lines are lines of constant $r$.
		The first panel~\ref{fig:A} shows $k=0$, the second~\ref{fig:B} shows $k=+1$, and the final panel~\ref{fig:C} shows $k=-1$, where the coordinates are bounded by $\tilde{t} > |x|$. }
	\label{fig1}
\end{figure}

In Figure~\ref{fig1} we plot the grid representing the coordinate transformation~(\ref{A_trans}) for the choice $R(t)=t$, for the three cases of $k=0$, $k=+1$ and $k=-1$. Here, we have allowed $r$ to be both positive and negative, due to the fact we have fixed the angular coordinates. The range of allowed values that $\tilde{t}$ and $x$ can take is clear from the plots, with the coordinates being unbounded for $k=0$ and $k>0$, or satisfying $\tilde{t} > \sqrt{|k|} |x|$ for negative $k$. For fixed values of $t$, the flat transformation $k=0$ also has fixed values of $\tilde{t}$, but for non-zero spatial curvature this is not the case. 
A fixed radius $r$ never corresponds to a fixed value of $x$, except for the origin with $r= x=0$. This gives some interpretation of what is happening during this coordinate transformation, and demonstrates the mixing of space and time in the coincident gauge system.
For more complicated choices of $R$, the axes are stretched but the general shape of the graph and the corresponding analysis remains the same. 

It is perhaps useful to see how these coordinates are inherently adapted to the Killing vectors. Take the $x$-translation generator in spherical coordinates with $\theta = \pi/2$ and $\phi=0$, 
\begin{equation*}
	\tau_{x} = \sqrt{1-k r^2} \partial_{r} \, .
\end{equation*}
Under the coordinate transformation this becomes
\begin{equation*}
	\hat{\tau}_{x} = - k r R  \partial_{\tilde{t}} + \sqrt{1-kr^2} R \partial_{x} \,  = -k x \partial_{\tilde{t}} + \tilde{t} \partial_{x} ,
\end{equation*}
as shown in Eq.~(\ref{tauk}) and~(\ref{taucoinck}). The coefficients of  $\partial_{\tilde{t}}$ and $\partial_{x}$ should be compared with~(\ref{A_trans}). Figure~\ref{fig1} then illustrates how the transformation brings the Killing vector into an affine form.

\clearpage 

\addcontentsline{toc}{section}{References}

\bibliographystyle{jhepmodstyle}
\bibliography{references}

\end{document}